\documentclass[aps,prl,twocolumn,reprint,preprintnumbers,nofootinbib,superscriptaddress,floatfix]{revtex4-2}
\usepackage{amsmath,amssymb}
\usepackage{graphicx}
\usepackage{booktabs}
\usepackage{xspace}
\usepackage{xcolor}
\usepackage{fontawesome5}
\usepackage[hidelinks]{hyperref}

\DeclareUnicodeCharacter{03B3}{\ensuremath{\gamma}}
\DeclareUnicodeCharacter{2192}{\ensuremath{\rightarrow}}
\let\inspiretextrightarrow\textrightarrow
\DeclareRobustCommand{\textrightarrow}{%
  \ifmmode\rightarrow\else\inspiretextrightarrow\fi}

\newcommand{\exhad}{\textsc{exHad}\xspace}
\newcommand{\mstar}{m_*}

\begin{document}
\interfootnotelinepenalty=10000

\preprint{CERN-TH-2026-223}

\title{Rethinking search signatures: Hadronic decays of GeV-scale feebly coupled particles}

\author{Viktor~Kryshtal}
\email{victor.kryshtal@gmail.com}
\affiliation{Department of Physics, Taras Shevchenko National University of Kyiv, 64 Volodymyrs'ka str., Kyiv 01601, Ukraine}
\author{Maksym~Ovchynnikov}
\email{maksym.ovchynnikov@cern.ch}
\affiliation{Theoretical Physics Department, CERN, 1211 Geneva 23, Switzerland}

\date{\today}

\begin{abstract}
Experimental searches for GeV-scale feebly interacting particles commonly
target the simplest decay signatures with two charged particles,
including hadrons. Hadronic modes often dominate the inclusive decay
rate, and their simulation relies extensively on \textsc{Pythia}.
Its fragmentation model, calibrated mainly on LEP data, is unreliable
in this mass range: it populates symmetry-forbidden states and
misidentifies the dominant allowed channels. We construct a simple
hadronization model tuned to electromagnetic scattering data and
constrained by conservation laws and available exclusive decay calculations, while retaining
a common prescription applicable to different particle models.
Considering the SHiP experiment as an example, we show that our model
shifts the dominant search signature after event selection from
two-particle to multiparticle decays, including mixed charged-neutral
states. We provide the model
in a form that can be readily integrated into \textsc{Pythia}-based
experimental simulation frameworks.
\end{abstract}

\maketitle

\paragraph{Introduction.}
Searches for GeV-scale feebly interacting particles target extensions
of the Standard Model such as heavy neutral leptons (HNLs), vector
mediators, including dark photons and vectors coupled to baryon minus
lepton number ($B-L$), axion-like particles (ALPs), and Higgs-like
scalars~\cite{Alekhin:2015byh,Beacham:2019nyx,Curtin:2018mvb}.
These particles can also mediate decays in richer dark sectors, including
heavier inelastic dark-matter states decaying through an off-shell dark
photon~\cite{Foguel:2024lca}.
Many existing searches and sensitivity studies for upcoming experiments
target final states containing only a few
particles, notably charged-lepton or meson pairs~\cite{Belle-II:2023ueh,
BaBar:2015jvu,Jaegle:2015fme,Belle:2020the,BABAR:2021cdg,
NA62:2025yzs,Craik:2022riw,MicroBooNE:2019izn,
SHiP:2018xqw,SHiP:2020vbd,FASER:2018eoc,Kholoimov:2025cqe,
Kholoimov:2025ycd}.

Decays into charged-lepton pairs provide the cleanest such signatures,
but their branching fractions are often strongly suppressed when
hadronic modes dominate~\cite{Aloni:2018vki,Ovchynnikov:2025gpx,Boiarska:2019jym,
DallaValleGarcia:2023xhh}.
This motivates hadronic searches and makes the probabilities for decays
into only a few hadrons crucial for sensitivity estimates.
For hadronic invariant masses above roughly $1$~GeV, the sum of the
calculated exclusive widths generally no longer accounts for the total
hadronic width. For hadronic systems with invariant masses above roughly 2~GeV,
the Lund string model implemented in \textsc{Pythia}~8 is commonly used to convert
quark- and gluon-level decays into mesons and other hadrons. Applications
include experimental detector simulations and sensitivity
studies~\cite{SHiP:2020vbd,SHiP:2018xqw,FASER:2018eoc,
Belle-II:2023ueh,LHCb:2025ymr,CMS:2024ake}, sensitivity projections
for accelerator-based experiments~\cite{Batell:2020vqn,Ovchynnikov:2023cry},
and calculations of cosmological hadronic
injection~\cite{Bianco:2025boy,
Bianco:2026dvc,Boyarsky:2020dzc,Akita:2026gee}.
The Lund fragmentation parameters were constrained mainly
by high-energy $e^+e^-$ data from LEP and SLD~\cite{Skands:2014pea}.
Applying this calibration to a complete hadronic system of a few GeV is
an extrapolation. We call fragmentation with the default parameters
raw \textsc{Pythia}.

\begin{figure*}[t!]
\centering
\resizebox{\textwidth}{!}{%
\includegraphics{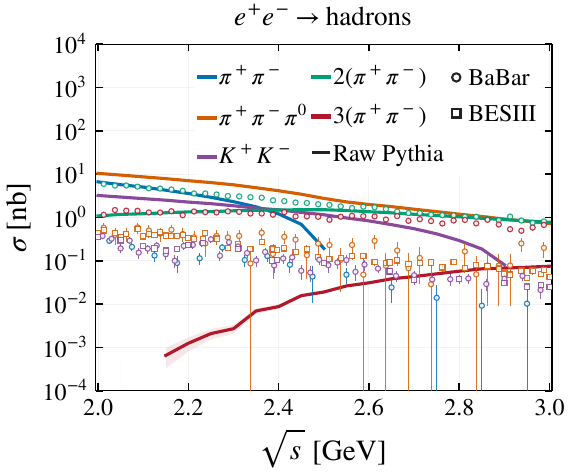}%
\includegraphics{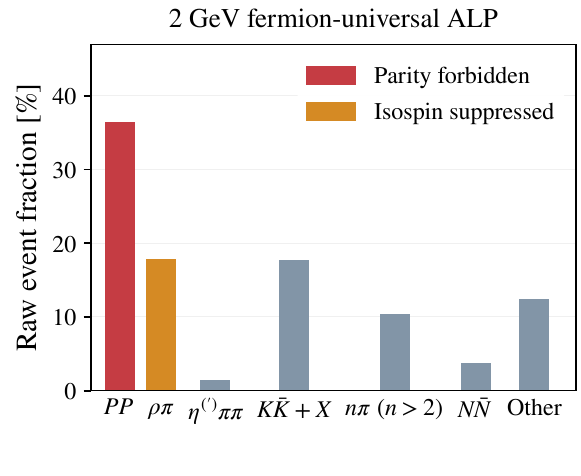}%
}
\caption{Failures of fragmentation in raw \textsc{Pythia} at low invariant mass of the hadronic system.
Left: electromagnetic cross sections from raw \textsc{Pythia} (solid lines),
BaBar (circles)~\cite{BaBar:2004ytv,
BaBar:2006vzy,BaBar:2012sxt,BaBar:2012bdw,BaBar:2013jqz} and
BESIII (squares)~\cite{BESIII:2018ldc,BESIII:2019gjz,BESIII:2024okl}.
Bands show Monte Carlo statistical errors. Right: hadronic decays of a
2~GeV ALP universally coupled to fermions at the scale $\Lambda=1$~TeV,
classified before resonance decays.
$PP$ denotes two pseudoscalar mesons and $N$ a nucleon.
The $K\bar K+X$ bin groups strange-hadron states, including primary
$K^*$ and $\phi$; $X$ denotes additional particles.
The $n\pi$ bin groups primary states made only of pions and $\rho$ mesons,
excluding the separately shown $\rho\pi$ bin.
$PP$ is parity-forbidden; $\rho\pi$ vanishes in the isospin limit.
The $\eta^{(\prime)}\pi\pi$ contribution is more than an order of
magnitude below the exclusive prediction~\cite{Balkin:2025enj,Ovchynnikov:2025gpx}.
Generator normalizations and cross-section conventions are specified in
SuM, Tab.~\ref{tab:raw-comparisons} and Sec.~\ref{sec:em-conventions}.
\label{fig:raw-failures}}
\end{figure*}

Fig.~\ref{fig:raw-failures} illustrates quantitative and qualitative
failures of the extrapolation. Raw \textsc{Pythia} produces
symmetry-forbidden states, misestimates allowed-channel rates by orders
of magnitude, and systematically overpredicts few-hadron final states.
This bias can misidentify the dominant signatures and lead experiments
to optimize their selections for the wrong ones.

Exclusive calculations~\cite{Aloni:2018vki,Ilten:2018crw,Boiarska:2019jym,
Ovchynnikov:2025gpx} cover a limited set of final states. Their summed
widths account for a rapidly decreasing fraction of the total inclusive
hadronic width as the particle mass increases.
For Higgs-mixed scalars, Ref.~\cite{Gieseke:2025gfq} estimates
branching fractions using a symmetry-constrained cluster model.
A study of multihadron HNL decays matched to quark-level rates appeared
just before this work~\cite{Schubert:2026mgf}.
The challenge is to extend this exclusive information with a fragmentation
model that respects the decaying particle's quantum numbers and predicts
the final-state composition.

\begin{figure*}[t!]
\centering
\resizebox{\textwidth}{!}{%
\includegraphics{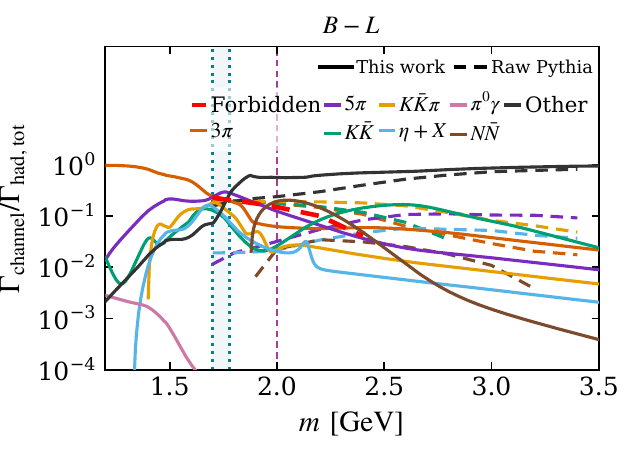}%
\includegraphics{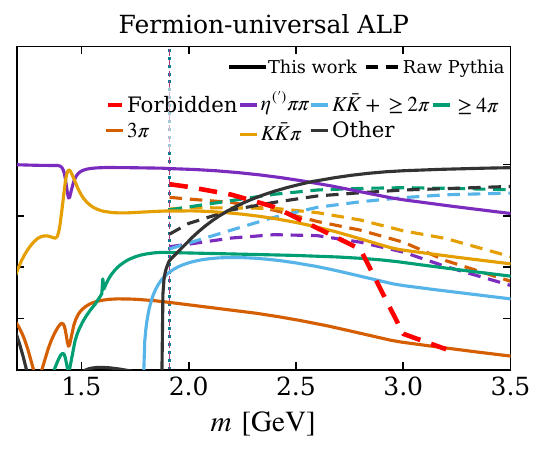}%
}
\caption{Hadronic channel fractions $\Gamma_i/\Gamma_{\rm had}$ for a
$B-L$ vector and a fermion-coupled ALP. Solid curves join exclusive
calculations to the tuned description; dashed curves show raw
\textsc{Pythia}. ``Forbidden'' (red dashed) curves show raw \textsc{Pythia}
fractions for $2\pi$ ($B-L$) and two-pseudoscalar states, including
$\pi\pi$, $K\bar K$, and $\eta\eta$ (ALP).
The $B-L$ current has no isovector ($\rho$-like) component, so $2\pi$
is absent in the isospin limit. The ALP two-pseudoscalar modes are
parity-forbidden.
Teal dotted lines bound total-width matching; purple dashed lines mark
the start of channel matching (SuM, Tab.~\ref{tab:matching-map}).
\label{fig:branchings}}
\end{figure*}

\begin{figure*}[t!]
\centering
\resizebox{\textwidth}{!}{%
\includegraphics{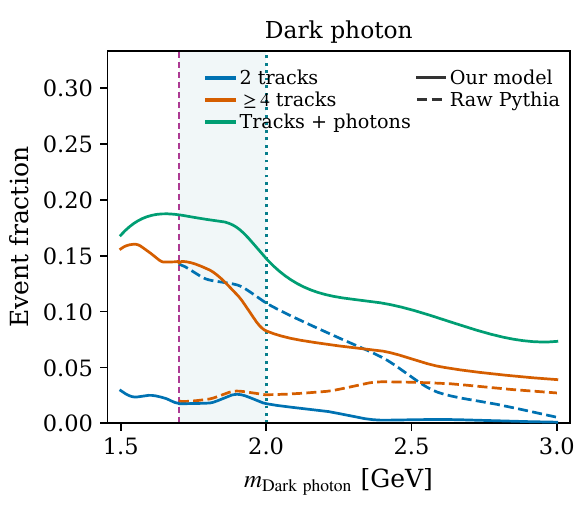}%
\includegraphics{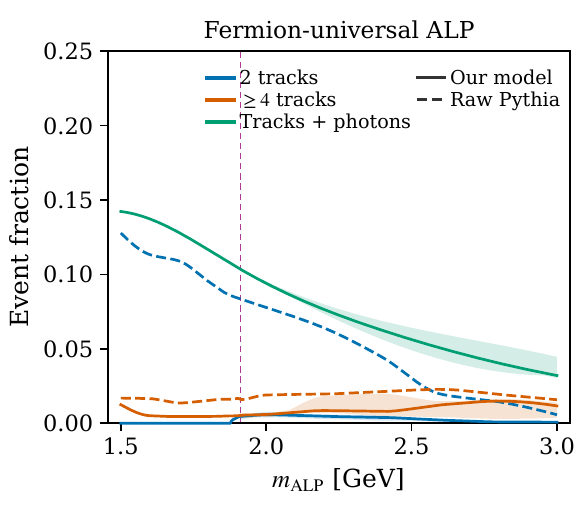}%
}
\caption{Accepted hadronic event fractions at SHiP, defined in
Eq.~\eqref{eq:ship-acceptance}. The panels show a dark photon (left) and
a fermion-coupled ALP (right), both with $c\tau=10$~m.
Blue and orange require exactly two or at least four final particles,
respectively, all charged. Green requires at least two charged particles
and at least one photon, with every final particle either charged or a photon.
Every particle must have laboratory momentum $|\mathbf p|>1$~GeV.
Charged particles must reach the spectrometer end and photons the
calorimeter end
(SuM, Sec.~\ref{sec:events-and-ship}).
Solid curves show our model; dashed blue and orange
curves use raw \textsc{Pythia}, shown for dark photons only at and above
the inclusive-switching mass of 1.7~GeV.
Curves are smooth guides to the calculated fractions; ALP bands show
systematic uncertainties in our hadronization model.
For dark photons, the change in curvature near $2$~GeV reflects the
changing production kinematics.
Purple dashed lines mark the start of the fragmentation contribution
at 1.7~GeV for dark photons and the start of channel-probability
matching at $\approx1.9$~GeV for ALPs. For dark photons, teal dotted
lines bound the 1.7--2~GeV interpolation of the total hadronic width.
\label{fig:ship}}
\end{figure*}

In this work, we develop a common prescription calibrated to
$e^+e^-\to\mathrm{hadrons}$ data and extended to other particles using
their exclusive rates and quantum numbers. For HNLs, we go beyond the
illustrative \textsc{Pythia} continuation of Ref.~\cite{Schubert:2026mgf}
by constraining multihadron compositions with hadronic data and
current-specific quantum-number restrictions. Our model favors
multihadron decays (Fig.~\ref{fig:branchings}), motivating multitrack and
mixed charged-neutral searches. Using SHiP as an example, we show that
mixed charged-photon states can dominate after event selection
(Fig.~\ref{fig:ship}).

We provide \exhad~\faGithub~\cite{exhad} as a standalone decay package
for \textsc{Pythia}-based simulations.
The Supplemental Material (SuM) details the prescription, inputs,
uncertainties, and implementation.

\paragraph{Lessons from electromagnetic data.}
A dark photon couples to the electromagnetic current, so measured
$e^+e^-\to\text{hadrons}$ cross sections determine its exclusive
hadronic widths relative to its muon-pair width~\cite{Ilten:2018crw}.
Decomposing the electromagnetic amplitude into its quark-current
components allows these measurements to constrain decay rates for
vectors with different quark couplings
\cite{Ilten:2018crw}. Vector-pseudoscalar production data also constrain
form factors for ALP decays to two vector mesons~\cite{Aloni:2018vki}.

Here and below, $s=q_{\rm had}^2$ is the squared invariant mass of the
complete hadronic system, with total four-momentum $q_{\rm had}$.
For fragmentation, this is also the invariant mass squared of the
initial quark or gluon system.
The measured cross sections for $e^+e^-\to2(\pi^+\pi^-)$ and
$K^+K^-\pi^+\pi^-$ decrease approximately as
$s^{-3}$~\cite{BESIII:2021ftf}. Relative to the inclusive light-quark
cross section proportional to $s^{-1}$, this motivates an $s^{-2}$
decrease in the probabilities for producing specified multihadron
final states, as in Eq.~\eqref{eq:model1}.
A possible explanation is production of two back-to-back
low-mass hadronic systems: producing additional hadrons within either
system need not make the cross section decrease more steeply with $s$.
A factorized four-pion calculation realizes this
$s^{-3}$ scaling~\cite{Bhattacharya:2025awq}.\footnote{Brodsky--Farrar
constituent-counting rules~\cite{Brodsky:1974vy,Lepage:1980fj} apply to
hard, fixed-angle configurations with well-separated hadrons.
Integrated multihadron cross sections also include collinear
configurations in which several hadrons form a low-mass system.
Their energy dependence requires accounting for these contributions,
so fixed-angle counting alone is insufficient to fix the power
(SuM, Sec.~\ref{sec:power-motivation}).}

\paragraph{A constrained fragmentation model.}
We reweight the predictions of \textsc{Pythia}'s Lund string model at low
hadronic invariant mass. This tuned model has three ingredients.
\emph{(i) Exclusive boundary conditions:} data or exclusive calculations
fix the channel probabilities at a reference mass $\mstar$.
\emph{(ii) Quantum-number conservation:} we reject generated hadronic
states that cannot carry the total isospin, $C$ parity, or $G$ parity of
the decaying particle's current; for two-body states we also require a
spin and parity compatible with the source.
\emph{(iii) Mass scaling:} Eq.~\eqref{eq:model1} extends the boundary
probabilities to larger invariant masses with a common $s^{-2}$ law,
modified near threshold by a finite-mass factor.

A channel group $F$ contains specified hadronic states whose total
probability is constrained. We classify these states after short-lived
resonances decay, retaining pions, kaons, $\eta$, $\eta'$, photons, and
nucleons. The detector signatures below include subsequent daughter decays.
We reweight channel probabilities while retaining \textsc{Pythia}'s momentum
distributions within each channel, targeting factor-of-two accuracy in exclusive
channel rates. Modes with separately calculated rates, such as two-meson
decays, are generated independently and excluded from fragmentation.

In the electromagnetic comparison, $3\pi$ denotes $\pi^+\pi^-\pi^0$,
and $4\pi$ combines $2(\pi^+\pi^-)$ and $\pi^+\pi^-2\pi^0$.
The group definitions, full model, and comparison with form-factor
extrapolations are given in SuM.
Among hadronic decays modeled by fragmentation, the probability of
producing a final state in channel group $F$ is
$P_F=\Gamma_F/\Gamma_{\rm frag}$. Here $\Gamma_F$ is the partial width
into that group, and $\Gamma_{\rm frag}$ is the hadronic width remaining
after subtracting the separately calculated channels.
Let $\mstar$ be the reference mass at which data or exclusive
calculations fix the channel probabilities. With $b_F=P_F(\mstar^2)$,
we use
\begin{equation}
 P_F(s)=\frac{b_F}{\mathcal N(s)}
        \left(\frac{\mstar^2}{s}\right)^p
        \frac{T_F(s)}{T_F(\mstar^2)},
 \qquad p\simeq2 .
 \label{eq:model1}
\end{equation}
Limited phase space suppresses the production of several hadrons near
threshold. The factor $T_F$ describes this turn-on and tends to a
constant at higher masses, leaving the data-motivated $s^{-2}$ falloff.
The ratio $T_F(s)/T_F(\mstar^2)$ preserves the input channel-group
probabilities at the reference mass. The factor $\mathcal N(s)$ keeps
the sum of the specified group probabilities at most one.
\textsc{Pythia} distributes the remaining probability among other
allowed channels. At each mass, our model and raw \textsc{Pythia} use
the same total hadronic width; our prescription redistributes this
width among decay channels. The common power is an empirical approximation
over a finite mass range; our central predictions use $p=2$.
The electromagnetic calibration uses $\mstar=2.32$~GeV, above the
prominent lower-mass vector peaks~\cite{BESIII:2022wxz}.

We extract $T_F$ from the generated channel probabilities after the
symmetry selection and exclusion of separately calculated modes.
We divide out each channel's fitted high-mass power law and retain a
rise larger than the Monte Carlo fluctuations; otherwise, $T_F=1$.

\paragraph{Comparison with electromagnetic data.}
The tuned model recovers the relative sizes of the measured three-,
four-, and six-pion cross sections (SuM, Fig.~\ref{fig:em-repair}).
Each channel group's rate and its division among charge combinations
are fixed at the reference mass. The common mass dependence then
describes 389 of 397 electromagnetic measurements within a factor of two.
This sample lies below the threshold for charmed-meson pair production
and excludes the narrow charmonium regions specified in SuM. The eight
remaining points are BaBar three-pion measurements between 2.3 and
3~GeV with uncertainties comparable to their values; they scatter on
both sides of the prediction. The finite-mass factor is especially important for six-pion
production compared with a pure power law. The full channel comparison
and broader data selections are given in SuM, Fig.~\ref{fig:em-repair}
and Tab.~\ref{tab:em-scores}; $K\bar K3\pi$ remains discrepant at higher
masses, but makes a very small contribution to the total hadronic cross
section. The quoted agreement is with the data used to develop the model.

\paragraph{From electromagnetic data to new particles.}
We extend the common prescription using each particle's exclusive decay
inputs and quantum-number restrictions; BC labels below denote the
benchmark cases of Ref.~\cite{Beacham:2019nyx}.
For decaying bosons, Eq.~\eqref{eq:model1} uses $s=m^2$, where $m$ is
the mass of the decaying particle.

For \textbf{dark photons}~\cite{Ilten:2018crw,Foguel:2022ppx}
(BC1~\cite{Beacham:2019nyx}),
electromagnetic data and exclusive calculations determine the rates of
many hadronic channels. The remaining channels
account for a growing fraction of the total hadronic width as the
dark-photon mass increases. \textsc{Pythia} generates decays into these
channels from 1.7~GeV. The total light-quark hadronic
width joins the sum of the calculated exclusive hadronic widths to the
perturbative inclusive width over
1.7--2~GeV. Channels with exclusive rates retain their separate
prescriptions for generating daughter momenta.
SuM, Sec.~\ref{sec:dark-photon-inputs}, specifies the matching
and the charge probabilities used for the unmeasured channels.

For the other particles, the matching masses follow the reach of the
exclusive calculations; omitted resonances remain a source of uncertainty.

For \textbf{$B-L$ vectors}~\cite{Ilten:2018crw,Foguel:2022ppx},
equal up- and down-quark couplings remove the isovector ($\rho$-like)
component of the current. Thus $2\pi$ is absent in the isospin limit,
while $3\pi$ is allowed. Electromagnetic form factors also give
substantial five-pion production through $\omega\pi\pi$.
Allowed two-body modes such as $K\bar K$ use
rates from data-anchored exclusive calculations.

For \textbf{fermion-coupled ALPs}~\cite{Aloni:2018vki,Ovchynnikov:2025gpx,Balkin:2025enj}
(BC10~\cite{Beacham:2019nyx}), existing hadronic-decay calculations
consistently favor
$\eta^{(\prime)}\pi\pi$ and $K\bar K\pi$ near the upper end of the
exclusive calculation, around 2~GeV. Their daughter decays
frequently contain neutral particles, while parity forbids two-pseudoscalar
final states. These features persist through matching to
Eq.~\eqref{eq:model1}.
For ALPs, this prototype hadronization model captures the qualitative
hierarchy of decay channels and its consequences for experimental searches.

\textbf{Higgs-like scalars}~\cite{Blackstone:2024ouf,Winkler:2018qyg,Boiarska:2019jym}
(BC4/BC5~\cite{Beacham:2019nyx}) permit two-pion and two-kaon decays. In the
exclusive input used here, four-pion modes account for the light-hadronic
width remaining after the separately calculated channels below 2~GeV.\footnote{We choose 2~GeV phenomenologically
as the boundary between the exclusive and fragmentation
descriptions~\cite{Winkler:2018qyg,Boiarska:2019jym}.}
SuM, Sec.~\ref{sec:scalar-inputs}, specifies the matching to fragmentation
and the continuation of the two-meson rates.

For \textbf{HNLs}~\cite{Bondarenko:2018ptm}
(BC6--BC8~\cite{Beacham:2019nyx}),
we apply the same prescription to multihadron decays induced
by charged or neutral weak currents involving $u$ and $d$ quarks.
Their channel probabilities use hadronic invariant-mass distributions
measured in $\tau$ decays and
$e^+e^-\to\mathrm{hadrons}$ form factors, with each current's own
normalization and symmetry restrictions~\cite{Davier:2013sfa,Ilten:2018crw}.
Eq.~\eqref{eq:model1} uses $s=W^2$, the squared hadronic invariant mass.
Integration over $W$ with the differential weak decay rate gives the
HNL channel fractions (SuM, Fig.~\ref{fig:sm-hnl-portals}).
This changes the composition while preserving the input one-meson rates,
total decay rates through each weak current, and HNL lifetime.

Fig.~\ref{fig:branchings} shows the fractions of the total hadronic
width carried by individual channels as functions of mass, comparing
our model with raw \textsc{Pythia} for a $B-L$ mediator and the same
fermion-coupled ALP as in Fig.~\ref{fig:raw-failures}.
For the matched multihadron channels, we join the exclusive inputs to
Eq.~\eqref{eq:model1} only over 2--3~GeV for $B-L$ and
$\approx1.9$--3~GeV for the ALP. Below these intervals, we use the
exclusive inputs; above them, we use Eq.~\eqref{eq:model1}.

Other models, such as gluon-coupled ALPs or inelastic dark matter,
can be described analogously.

\paragraph{Consequences for searches.}
The shift toward multihadron decays, illustrated in
Fig.~\ref{fig:branchings}, changes the relevant search signatures across
particle models, including dark photons and Higgs-like scalars.
These decays often contain both charged and neutral particles.
Selections restricted to charged-pair final states can reject multihadron
decays. For a 2-GeV ALP, the
$\eta^{(\prime)}\pi\pi$ modes account for about 80\% of its hadronic width in
Fig.~\ref{fig:branchings}. These modes include
$a\to\eta\pi^+\pi^-$ followed by $\eta\to\gamma\gamma$, giving
two tracks and two photons. The charged tracks can reveal a displaced
decay vertex; reconstructing the parent mass also requires calorimetry
for the photons. The $B-L$ $3\pi$ mode similarly gives
$\pi^+\pi^-\pi^0$, followed by $\pi^0\to\gamma\gamma$.

These changes in decay composition affect searches at many experiments,
including LHCb~\cite{LHCb:2025ymr,Kholoimov:2025cqe,Kholoimov:2025ycd,Gorkavenko:2023nbk}, ATLAS~\cite{ATLAS:2025pak},
CMS~\cite{CMS:2024ake}, DarkQuest~\cite{Batell:2020vqn}, and
SHiP~\cite{SHiP:2020vbd}, as well as
proposed experiments such as ANUBIS~\cite{ANUBIS:2025sgg},
MATHUSLA~\cite{Curtin:2023skh}, GRENDEL~\cite{Citron:2026kgx},
CODEX-b~\cite{CODEX-b:2025rck}, and those at the Forward Physics
Facility (FPF)~\cite{Feng:2022inv}.

We consider SHiP as a particular example. Most signal studies there
rely on reconstructing a charged pair, as
in dark-photon and HNL searches~\cite{SHiP:2020vbd,Graverini:2214085,SHiP:2018xqw,Bahmani:2025fqd}.
Calorimetric reconstruction of purely photonic decays has also been
investigated~\cite{Climescu:2026esp}. To our knowledge, no dedicated SHiP
study has yet established the full-event reconstruction performance
for mixed charged-neutral decays.

Two-body decays have a geometric advantage in a forward detector:
fewer daughter trajectories must fall inside its coverage. To test
whether this compensates for their reduced probabilities, we compare
dark-photon and ALP decays generated with \exhad in
\textsc{EventCalc-SHiP}~\cite{EventCalc} at 1.5--3~GeV and $c\tau=10$~m.
For a category $\mathcal C$
defined by final-particle content after daughter decays, we calculate
the accepted fraction
\begin{equation}
f_{\mathcal C}^{\mathrm{rec}} =
\frac{N_{\mathcal C}^{\text{all }n\text{ particles accepted}}}
{N_{\mathrm{had}}^{\mathrm{in\ volume}}}.
\label{eq:ship-acceptance}
\end{equation}

The numerator requires all $n$ final particles to pass the geometric
selection and laboratory momentum cut $|\mathbf p|>1$~GeV, including
photons~\cite{Albanese:2878604}; the denominator counts all hadronic decays
inside the volume.
Both include production and decay-probability weights.
This fraction combines the probability of producing the selected final
state with the acceptance for all its decay products.
Fig.~\ref{fig:ship} defines the three disjoint categories; SuM,
Sec.~\ref{sec:events-and-ship}, gives the geometry and simulation details.
We assume unit efficiency for reconstructing each accepted event in
full. Detector-level simulations must establish the reconstruction
efficiency for multitrack and mixed charged-neutral events; the
straw-tracker study of Ref.~\cite{Lelak:2024NA62} suggests an
$\mathcal O(1)$ efficiency for charged-only states.

In the 2--3~GeV region, our model suppresses two-track
dark-photon decays and enhances the higher-multiplicity all-charged
contribution. At 2~GeV, the accepted fraction for fully charged decays
with at least four particles divided by the accepted fraction with
exactly two is about $1/4$ in raw \textsc{Pythia} and five in our model.
The dominant fully charged category therefore reverses despite the
suppression from geometric acceptance.

The ALP bands in Fig.~\ref{fig:ship} show the spread from variations of
the channel-probability law and \textsc{Pythia} fragmentation in the
charged-only and mixed selections, with the exclusive inputs and matching choices
held fixed (SuM, Sec.~\ref{sec:uncertainties}).

The momentum generator also affects acceptance: for the same ALP
$\eta\pi^+\pi^-$ channel and parent events, replacing the exclusive
matrix element with \textsc{Pythia} momenta changes the mixed charged-photon
acceptance by 5--6\%. Separate dark-photon samples on either side of
1.7~GeV differ by approximately 2\% in their mixed accepted fraction
(SuM, Sec.~\ref{sec:ship-momentum-boundary}).

For the ALP, the accepted fully charged two-particle fraction at 2~GeV
is about fourteen times smaller than the raw-\textsc{Pythia} prediction;
the suppression exceeds an order of magnitude for every tested
hadronization-model variation at this mass. The accepted fraction with
at least four particles, all charged, is also suppressed at this mass
across the tested variations.

Mixed charged-photon states give the largest accepted fraction in both panels.
At 2~GeV, adding this category increases the combined accepted fraction
by about a factor of 2.4 for dark photons and nine for ALPs relative
to the two charged-only categories.

% Keep normal paragraph spacing if the closing notices move to the next page.
\raggedbottom
\paragraph{Conclusions.}
We have developed a common model of hadronic decays that combines
particle-specific exclusive rates and quantum-number restrictions with a
common empirical mass dependence constrained by electromagnetic data.
Relative to raw
\textsc{Pythia}, our results favor multihadron final states, often
containing neutral particles, over charged-hadron pairs. These changes
in decay composition have implications across experiments searching for
GeV-scale feebly coupled particles. Our SHiP examples demonstrate that
they can change the dominant search signatures even after geometric and
momentum selections. The results motivate multitrack searches,
reconstruction using both charged tracks and photons, and a reassessment
of sensitivities based on default fragmentation. A closer fit to
electromagnetic data alone does not establish higher accuracy for other
particles. Establishing such accuracy requires control of quantum-number
correlations in hadron formation, missing resonances, interference between
amplitudes producing the same final state, and correlations among daughter
momenta. The model can incorporate improved exclusive descriptions.
The \exhad implementation~\cite{exhad} provides these decay
descriptions for \textsc{Pythia}-based simulations.

\section*{Acknowledgements}
We thank Stephen Mrenna for reading the main text and providing ideas
and technical input for this study.
We thank Philip Ilten for useful discussions, and Miguel Escudero and
Arantza Oyanguren for reading the manuscript. We thank Matei Climescu
for reading the draft and for discussions on signal reconstruction at
SHiP. MO thanks Fermilab for its
hospitality during a summer research visit. We used Astra 6 and Fable 5.1
for technical assistance with software,
numerical checks, and language editing. The authors remain
responsible for the scientific content.

\section*{Conflict of interest statement}
MO is a member of the SHiP collaboration.
The views expressed in this manuscript are those of the authors alone.

\bibliographystyle{apsrev4-2-inspire}
% REVTeX skips its bibliography control entry (article titles, eprints) when
% \bibliographystyle is given explicitly; restore that behavior.
\makeatletter\@booleanfalse\bibliographystyle@sw\makeatother
\bibliography{refs}

\onecolumngrid
% The Supplemental Material starts on a separate page.
\clearpage
\flushbottom
\pdfbookmark[0]{Supplemental Material}{supplemental-material}
\setcounter{section}{0}
\setcounter{equation}{0}
\setcounter{figure}{0}
\setcounter{table}{0}
\setcounter{secnumdepth}{2}
\renewcommand{\thesection}{S\Roman{section}}
\renewcommand{\theequation}{S\arabic{equation}}
\renewcommand{\thefigure}{S\arabic{figure}}
\renewcommand{\thetable}{S\Roman{table}}
\renewcommand{\theHsection}{supplement.\arabic{section}}
\renewcommand{\theHequation}{supplement.\arabic{equation}}
\renewcommand{\theHfigure}{supplement.\arabic{figure}}
\renewcommand{\theHtable}{supplement.\arabic{table}}
\begin{center}
{\large\bfseries Supplemental Material}\par\medskip
{\bfseries Rethinking search signatures:\\
Hadronic decays of GeV-scale feebly coupled particles}
\end{center}
\medskip
\fontsize{11}{13.2}\selectfont
Sec.~\ref{sec:conditional-model} defines the hadronic channel probabilities,
their mass dependence, matching to exclusive calculations, and event generation.
Sec.~\ref{sec:currents} compares the model with electromagnetic data.
Sec.~\ref{sec:portal-inputs} describes the particle-specific inputs and matching prescriptions.
Sec.~\ref{sec:decay-results} interprets the channel fractions, their
uncertainties, and the accepted signatures at SHiP.
Sec.~\ref{sec:implementation} describes the interface, sampling,
and numerical implementation.

\section{Common decay prescription}
\label{sec:conditional-model}

The prescription specifies how often each channel occurs and how its
particles share the energy. It uses the decaying particle's currents and
conservation laws, exclusive rates at a reference mass, and the
high-mass dependence motivated by electromagnetic data.
\textsc{Pythia} generates hadronic events with model probabilities for groups of
channels. We call this prescription the tuned model.
For dark photons, our model uses the measured electromagnetic rates
where available, as described in Sec.~\ref{sec:dark-photon-inputs}.

\subsection{Channel definitions and normalization}
We generate $u\bar u$, $d\bar d$, $s\bar s$, or $gg$ sources separately.
A source specifies both the initial quark-antiquark or gluon pair
and the quantum numbers of the current producing it.
For a vector current, the isovector light, isoscalar light, and
isoscalar strange components are denoted $\rho$-like, $\omega$-like,
and $\phi$-like. The first two use equal $u\bar u$ and $d\bar d$
samples; the last uses $s\bar s$. We reject primary hadronic states that
violate the current's symmetry restrictions. String breaking can create $s\bar s$ within
a light-quark string, so a light isoscalar source can produce kaons.

We classify each event after decaying the prompt resonances, leaving
pions, kaons, $\eta$, $\eta'$, photons, and nucleons undecayed at this
stage. Subsequent daughter decays are included when constructing the
experimental signatures.
The resulting hadronic state must satisfy the charge, strangeness,
total isospin, charge-conjugation parity ($C$), and $G$-parity ($G$)
restrictions of the current, listed in Tab.~\ref{tab:symmetry-constraints}.
String fragmentation conserves charge and net strangeness automatically.
Total isospin, $C$, and $G$ are not conserved by \textsc{Pythia}; we
impose them by rejecting generated states that violate them. For
two-body states we also require a spin and parity compatible with the
source.
A \emph{channel group} $F$ contains the specified final states assigned
one common probability correction. Tab.~\ref{tab:em-groups} gives the
electromagnetic groups. Other particles have different groups fixed by
their exclusive calculation.

\begin{table}[!htb]
\centering
\begin{tabular}{lcccc}
\toprule
Source & $J^P$ & $I,\,I_3$ & $G$ & $C$\\
\midrule
Dark photon, $\rho$-like & $1^-$ & $1,\,0$ & $+$ & $-$\\
Dark photon and $B-L$, $\omega$-like and $\phi$-like & $1^-$ & $0,\,0$ & $-$ & $-$\\
Fermion-coupled ALP, $gg$ and $s\bar s$ & $0^-$ & $0,\,0$ & $+$ & $+$\\
Higgs-like scalar, $gg$ and $s\bar s$ & $0^+$ & $0,\,0$ & $+$ & $+$\\
HNL $CC_{ud}$, vector & $1^-$ & $1,\,\pm1$ & $+$ & ---\\
HNL $CC_{ud}$, axial & $1^+$ & $1,\,\pm1$ & $-$ & ---\\
HNL $NC_{ud}$, isovector vector & $1^-$ & $1,\,0$ & $+$ & ---\\
HNL $NC_{ud}$, isovector axial & $1^+$ & $1,\,0$ & $-$ & ---\\
HNL $NC_{ud}$, isoscalar vector & $1^-$ & $0,\,0$ & $-$ & ---\\
\bottomrule
\end{tabular}
\caption{Quantum numbers imposed on generated hadronic states for each
source. $CC_{ud}$ and $NC_{ud}$ denote the charged and neutral weak
currents in the nonstrange $u,d$ sector (Sec.~\ref{sec:hnl-inputs}).
\textsc{Pythia} conserves charge and net flavor by itself. A
generated state is rejected if its hadron isospins cannot couple to the
listed $I$ and $I_3$; if it consists of integer-isospin mesons with
defined $G$ parity and their product differs from the listed $G$ (states
with baryons or kaons are not tested); if it consists only of
self-conjugate neutral mesons and the product of their $C$ parities
differs from the listed $C$; or if it is a two-body state with no orbital
and spin combination giving $J^P$. Charged currents have no $C$ parity;
for the neutral current, $C$ follows from $G$ and $I$ and is not imposed
separately. The $I_3$ sign of the charged current follows the lepton
charge.}
\label{tab:symmetry-constraints}
\end{table}

For example, $\omega\pi^+\pi^-$ followed by
$\omega\to\pi^+\pi^-\pi^0$ enters the five-pion group.
An $\eta\omega$ state enters the $\eta3\pi$ group used for $B-L$.
Its $\eta$ subsequently decays when the complete event is prepared for
detector simulation. The correction is determined before that decay.

\paragraph{Rates and their normalization.}
Let $\Gamma_{\rm tot}$ be the full decay width and $\Gamma_{\rm had}$
the width into states containing hadrons, including hadrons with photons.
Purely leptonic, invisible, and diphoton decays contribute to
$\Gamma_{\rm tot}$ separately. Let $\Gamma_{\rm sep}$ be the sum of
hadronic widths calculated and generated separately, including heavy
flavors where present. Fragmentation describes
$\Gamma_{\rm frag}=\Gamma_{\rm had}-\Gamma_{\rm sep}$.
For $s=m^2$, let $\Gamma_F$ be the partial width into group $F$ and
$R$ the allowed states outside the explicitly constrained groups.
Their probabilities within fragmentation are
\begin{equation}
 P_F(s)=\frac{\Gamma_F(s)}{\Gamma_{\rm frag}(s)},\qquad
 P_R(s)=1-\sum_F P_F(s).
 \label{eq:conditional-definition}
\end{equation}
The boson plots show
$\Gamma_F/\Gamma_{\rm had}=(\Gamma_{\rm frag}/\Gamma_{\rm had})P_F$;
the full branching ratio is
$\mathrm{Br}_F=(\Gamma_{\rm frag}/\Gamma_{\rm tot})P_F$.
Separate channels use their own partial widths in these ratios.
For non-electromagnetic currents, we use the same functional form for
the channel-group probabilities in Eq.~\eqref{eq:portable-law}, with
inputs determined separately for each current. Partial widths follow from
$\Gamma_F(s)=P_F(s)\Gamma_{\rm frag}(s)$, so their mass dependence also
contains the current-specific fragmentation width.
A plotted Other category may combine $R$ with calculated channels that
have no individual curve; its contents are specified for each figure.

\begin{table*}[t!]
\centering
\begin{tabular}{p{0.23\textwidth}p{0.66\textwidth}}
\toprule
Group & Hadronic states included\\
\midrule
$3\pi$ & $\pi^+\pi^-\pi^0$\\
$4\pi$ & $2(\pi^+\pi^-)$, $\pi^+\pi^-2\pi^0$\\
$5\pi$ & $2(\pi^+\pi^-)\pi^0$\\
$6\pi$ & $3(\pi^+\pi^-)$, $2(\pi^+\pi^-)2\pi^0$\\
$\eta\pi\pi$ & $\eta\pi^+\pi^-$\\
$\eta'\pi\pi$ & $\eta'\pi^+\pi^-$\\
$K\bar K\pi$ & $K^+K^-\pi^0$, $K^+\bar K^0\pi^-$ and its charge
conjugate, $K^0\bar K^0\pi^0$ (represented also as $K_SK_L\pi^0$)\\
$K\bar K2\pi$ & $K^+K^-\pi^+\pi^-$, $K^+K^-2\pi^0$\\
$K\bar K3\pi$ & $K^+K^-\pi^+\pi^-\pi^0$\\
$4K$ & $2(K^+K^-)$\\
$R$ & All other allowed generated states, excluding separately calculated channels\\
\bottomrule
\end{tabular}
\caption{The ten channel groups used for the electromagnetic comparison,
with probabilities normalized as in Eq.~\eqref{eq:conditional-definition}.
Only the listed charge states belong to these groups. An additional
allowed particle or a different unlisted charge combination places a
state in $R$, after removing channels calculated separately.
$\eta$ and $\eta'$ are counted before their later decays.}
\label{tab:em-groups}
\end{table*}

\subsection{Mass dependence of channel probabilities}
\label{sec:power-motivation}
At the reference mass $\mstar$, exclusive rates fix
$b_F=P_F(\mstar^2)$. Restating Eq.~\eqref{eq:model1} from the main text
with the normalization and remainder made explicit, we first define
the group weights $u_F$ before applying the normalization $\mathcal N(s)$:
\begin{equation}
 \begin{aligned}
 u_F(s)&=b_F\left(\frac{\mstar^2}{s}\right)^p
 \frac{T_F(s)}{T_F(\mstar^2)},\qquad p=2,\\
 \mathcal N(s)&=\max\left[1,\sum_F u_F(s)\right],\qquad
 P_F(s)=\frac{u_F(s)}{\mathcal N(s)},\qquad P_R(s)=1-\sum_F P_F(s).
 \end{aligned}
 \label{eq:portable-law}
\end{equation}
Here, $T_F$ describes a rise in the ratio obtained by dividing the
\textsc{Pythia} channel probability by its fitted power-law decrease.
We retain this rise only when it exceeds Monte Carlo fluctuations and
approaches a constant at larger mass.
Sec.~\ref{sec:finite-mass} defines its extraction.
The probability-normalization factor $\mathcal N(s)$ rescales all groups
together if their sum would exceed one.
The particle-specific matching prescriptions are given in
Sec.~\ref{sec:portal-inputs} and Tab.~\ref{tab:matching-map}.

The exponent $p$ controls how rapidly the probability of a specified
multihadron group falls with mass: $P_F\propto s^{-p}$ once the
finite-mass factor has saturated and $\mathcal N(s)=1$ in Eq.~\eqref{eq:portable-law}.
For an electromagnetic fragmentation cross section proportional to
$s^{-1}$, this gives $\sigma_F\propto s^{-(p+1)}$; our central choice
$p=2$ therefore corresponds to $\sigma_F\propto s^{-3}$.
BESIII obtains $\sigma\propto s^{-3.12}$ for
$K^+K^-\pi^+\pi^-$ and $s^{-3.18}$ for $2\pi^+2\pi^-$ at
hadronic invariant masses $\sqrt{s}$ from approximately 3.77 to
4.6~GeV~\cite{BESIII:2021ftf}.
With an inclusive cross section proportional to $s^{-1}$, these slopes correspond to
$p\simeq2.12$ and $2.18$. The central model uses the simple common value
$p=2$; we vary this common exponent from 1.75 to 2.25 simultaneously
for all specified groups of a particle. The measured powers vary across channels: the
same analysis gives $\sigma\propto s^{-2.49}$ for
$K^+K^-\pi^+\pi^-\pi^0$. Thus this variation tests the common-power
assumption; its range has no statistical confidence-level interpretation.

Constituent-counting rules assign different powers to final states of
different multiplicity when every hadron participates in a hard fixed-angle
configuration~\cite{Brodsky:1974vy,Lepage:1980fj}. An integrated multihadron
cross section also contains configurations with several nearby hadrons.
For example, Ref.~\cite{Bhattacharya:2025awq} calculates
$e^+e^-\to(\pi^+\pi^0)(\pi^-\pi^0)$ when both dipion invariant masses
remain small compared with $s$. Eq.~(37) of that reference has an
$s^{-3}$ leading power, up to running and evolution effects. Integration
over fixed low dipion-mass intervals retains this power. Here the two
pions in each pair are nearly parallel, and the two pairs move
back-to-back. Fixed-angle counting for four separate mesons assumes
large angles between every pair of mesons. The small opening angles
within the pion pairs violate that assumption.

The two low-invariant-mass hadronic systems in this example motivate
testing a common power for multihadron production from a quark or gluon
pair. We apply it to integrated channel probabilities as an empirical
prescription, tested against electromagnetic data. For each new current,
we determine its quark/gluon source weights, symmetry restrictions, reference
probabilities, and $T_F$, while retaining the same exponent.

\paragraph{Relation to electromagnetic form-factor extrapolations.}
Ref.~\cite{Ilten:2018crw} obtains other vector-current widths by flavor
decomposition of electromagnetic data. Ref.~\cite{Aloni:2018vki} extracts
a dimensionless mass-dependent factor $\mathcal F(m)\propto m^{-4}$ above
2~GeV, multiplying the vector-meson-dominance $VVP$ amplitude; $V$ and
$P$ denote vector and pseudoscalar mesons. The amplitude's Lorentz
structure contains two explicit momenta. At fixed daughter masses,
$\Gamma_{V\to VP}\propto m^3|\mathcal F|^2$, and Eq.~(16) of that
reference gives $\sigma_{VP}\propto\Gamma_{V\to VP}/m^3\propto s^{-4}$
asymptotically. Converting this amplitude factor into a channel
probability also requires the momentum factors, phase space, and
normalization to the inclusive rate.

The electromagnetic measurements used in these approaches overlap,
but the contributions being extrapolated need not coincide.
Ref.~\cite{Aloni:2018vki} uses the $\omega\pi^0$ subchannel extracted
from $e^+e^-\to\pi^+\pi^-2\pi^0$, whereas our four-pion group includes
both $2(\pi^+\pi^-)$ and $\pi^+\pi^-2\pi^0$. BaBar finds that
$\omega\pi^0$ accounts for about 10\% of the $\pi^+\pi^-2\pi^0$ rate at
1.8~GeV, with a decreasing fraction above it~\cite{BaBar:2017zmc}.
The other contributions include resonant processes, leaving the relative
importance of direct, nonresonant four-pion production unresolved by this comparison.

Thus the total four-pion rate and its $\omega\pi$ contribution need not
have the same power. The three-pion comparison is different: the input of
Ref.~\cite{Aloni:2018vki} is the total measured
$\pi^+\pi^-\pi^0$ rate interpreted through $\rho\pi$.
For such a $VP$-dominated final state, the $s^{-3}$ and $s^{-4}$
continuations are genuinely different approximations. If normalized at
the same mass $\mstar$, their ratio grows as $s/\mstar^2$, giving
about 1.7 at 3~GeV for our reference mass. This modest difference over
a limited interval can lie within the factor-of-two accuracy of the
model. Establishing their asymptotic behavior or a crossover to dominance
by other amplitudes requires further analysis.

\subsection{Finite-mass correction}
\label{sec:finite-mass}

The energy available after producing the hadron masses strongly affects
the probability of a many-particle final state. \textsc{Pythia} also introduces
mass dependence through string termination and its fragmentation
parameters~\cite{Bierlich:2022pfr}. Dividing out the fitted overall
power-law fall leaves the generator's remaining mass dependence.
We retain only a statistically significant rise that approaches a plateau.
Using this residual shape as a finite-mass correction is a phenomenological
assumption, assessed against electromagnetic data.
The correction therefore changes the approach to the common power law
without changing its high-mass exponent.

The resulting factors matter most for higher multiplicities
(Tab.~\ref{tab:finite-mass-plateaus}). Five electromagnetic groups
need no correction to the pure probability power. For six pions, the
$T_F$ in Eq.~\eqref{eq:portable-law} more than doubles the high-mass probability relative
to the pure power law normalized at the reference mass and recovers the agreement with
data lost in that approximation (Tab.~\ref{tab:em-scores}). The factor
combines the effects of phase space and string fragmentation; the
electromagnetic comparison tests their net effect on the probabilities.

We extract the factor from the channel probability $Q_F^{\rm Py}$
with the particle's source weights and channel selection. Its reference
mass $m_0$ fixes the normalization of the generator comparison;
$\mstar$ fixes the physical probability from data or an exclusive
calculation. Tab.~\ref{tab:generator-reference-masses} gives both.
With $s_0=m_0^2$, the extraction has three steps.

\begin{enumerate}
\item Fit $Q_F^{\rm Py}(s)=A_Fs^{-p_F^{\rm Py}}$ over
$3\leq\sqrt{s}\leq5$~GeV, with fitted normalization $A_F$ and
exponent $p_F^{\rm Py}$, and divide out this power:
\begin{equation}
 C_F^{\rm Py}(s)=
 \frac{Q_F^{\rm Py}(s)}{Q_F^{\rm Py}(s_0)}
 \left(\frac{s}{s_0}\right)^{p_F^{\rm Py}}.
 \label{eq:py-residual}
\end{equation}
The ratio $C_F^{\rm Py}$ equals one at $s_0$ and is constant for a pure power.
The fitted high-mass limit is
$L_F=A_F/[Q_F^{\rm Py}(s_0)s_0^{p_F^{\rm Py}}]$.

\item Accept the rise only if the fit of $\ln Q_F^{\rm Py}$ against
$\ln s$ has coefficient of determination $R^2\geq0.95$; $L_F>1$;
$\ln C_F^{\rm Py}$ at the first point above
$m_0$ is positive by more than five estimated standard deviations;
and $|\ln(C_F^{\rm Py}/L_F)|<0.10$ at every sampled mass at or above
3.5~GeV. Otherwise, set $T_F=1$.

\item After dividing out the fitted power, the \textsc{Pythia}
probability of a high-multiplicity group rises from one at $m_0$ and
levels off at $L_F$. This rise has statistical noise. For groups that
pass these tests, we keep only the rise: at each sampled squared mass
$s_i$, in ascending order, we take the largest value of $C_F^{\rm Py}$
reached so far, but not above $L_F$:
\begin{equation}
 T_F(s_i)=\min\left[L_F,\,
             \max\left(1,\max_{j\leq i}C_F^{\rm Py}(s_j)\right)\right].
 \label{eq:finite-mass-factor}
\end{equation}
The factor $T_F$ therefore grows from one to $L_F$ and never decreases.
A monotone cubic interpolation of $\ln T_F$ against $\ln s$ connects
these nodes. Set $T_F=1$ below $m_0$ wherever the probability
continuation requires it.
\end{enumerate}

The probability estimators and statistical errors used in these tests
are specified in Sec.~\ref{sec:implementation}.

Below 3~GeV we use only the shape of the generator's mass dependence,
normalized to one at $m_0$; the level is fixed by data at $\mstar$. The
retained rise models the limited energy available for many hadrons near
threshold. Two safeguards limit what is taken from the
generator. The gates in step 2 discard rises that are not statistically
resolved or that do not level off. The reference mass $m_0$ lies above
the prominent vector resonances; an exploratory use of the full residual
from 2~GeV gave poorer agreement (Sec.~\ref{sec:em-comparison-definition}).
The six-pion data test the imported shape directly: with $T_F$, 98 of 98
six-pion points fall within a factor of two, compared with 51 of 98 for
the pure power (Tab.~\ref{tab:em-scores}).

\begin{table*}[t!]
\centering
\begin{tabular}{lcc}
\toprule
Application & Generator $m_0$ [GeV] & Probability reference $\mstar$ [GeV]\\
\midrule
Electromagnetic test & 2.32 & 2.32\\
Fermion-coupled ALP & $\approx1.9$ & $\approx1.9$\\
Higgs-like scalar & 2 & 2\\
$B-L$ & 2 & 3\\
HNL charged current & 1.651 & 1.65\\
HNL neutral current & 3 & 3\\
\bottomrule
\end{tabular}
\caption{Reference masses for the factor $T_F$, defined in
Eq.~\eqref{eq:finite-mass-factor}, and the probability law,
Eq.~\eqref{eq:portable-law}. For HNLs, all entries refer to the hadronic
mass $W$. The $B-L$ multihadron partial widths are joined on 2-3~GeV;
Eq.~\eqref{eq:portable-law} is normalized at 3~GeV using $T_F(m^2)/T_F(9~\mathrm{GeV}^2)$.}
\label{tab:generator-reference-masses}
\end{table*}

\begin{table*}[t!]
\centering
\begin{tabular}{lc}
\toprule
Channel group & $T_F(\text{high mass})$\\
\midrule
$3\pi$, $\eta\pi\pi$, $\eta'\pi\pi$, $4\pi$, $K\bar K\pi$ & 1\\
$5\pi$ & 1.38\\
$6\pi$ & 2.37\\
$K\bar K2\pi$ & 1.20\\
$K\bar K\pi\pi\pi^0$ & 3.33\\
$4K$ & 3.36\\
\bottomrule
\end{tabular}
\caption{High-mass values of $T_F$, defined in Eq.~\eqref{eq:finite-mass-factor}
and multiplying the probability in Eq.~\eqref{eq:portable-law},
for the electromagnetic comparison.
Each $T_F$ starts at one at 2.32~GeV. Unit entries use $T_F=1$
throughout.}
\label{tab:finite-mass-plateaus}
\end{table*}

\subsection{Matching to exclusive calculations}
\label{sec:alp-probability-matching}

Near the end of an exclusive calculation, the decay probabilities should
connect smoothly to their high-mass continuation.
The reference mass $\mstar$ normalizes the probabilities. The start of
\textsc{Pythia} generation is specified separately. We distinguish three operations:
joining the channel probabilities, joining the total hadronic width,
and changing the generator that produces the hadronic events.
Tab.~\ref{tab:matching-map} specifies the quantities and intervals used
for each particle. Matching is confined to the stated intervals; the
exclusive inputs and high-mass continuation determine the mass dependence
on either side. The channel-matching polynomials for dark photons,
$B-L$, and ALPs reproduce the value and mass derivative of the adjacent
description at each endpoint. The $B-L$ total-width matching also
reproduces the second derivative. The scalar begins its continuation directly, without
a separate slope-matching interval.

Continuity of channel probabilities does not by itself guarantee
continuous absolute widths: these also depend on $\Gamma_{\rm frag}$,
and hence on the total width and the separately calculated rates.
The particle-specific constructions are given in Sec.~\ref{sec:portal-inputs}.

\subsection{Event generation}
\label{sec:event-generation}
For source $a$, let $Q^{\rm Py}_{F,a}$ be the \textsc{Pythia} probability of
group $F$ after the current restrictions and removal of separately
calculated channels. Let $r_a$ be the relative probability of that source,
determined from the particle's couplings and partonic widths.
The probability after combining the quark and gluon sources is
\begin{equation}
 Q_F^{\rm Py}(s)=\sum_a r_a(s)Q^{\rm Py}_{F,a}(s),\qquad
 \sum_a r_a(s)=1.
 \label{eq:source-mixture}
\end{equation}

Let $Q_R^{\rm Py}=1-\sum_F Q_F^{\rm Py}$ be the generated remainder
probability. An event in group $F$ or $R$ receives the respective weight
\begin{equation}
 w_F(s)=\frac{P_F(s)}{Q_F^{\rm Py}(s)},\qquad
 w_R(s)=\frac{P_R(s)}{Q_R^{\rm Py}(s)}.
 \label{eq:weights}
\end{equation}
By themselves, these group weights change the relative probabilities of
groups while retaining their generated momenta and charge composition.
The power fit uses
$Q_F^{\rm Py}$ directly, with all fragmentation channels in its denominator.

When data, an exclusive calculation, or isospin specify the relative rates of
different charge combinations within a group, let $f_{i|F}$ and
$q_{i|F}^{\rm Py}$ denote the model and \textsc{Pythia} probabilities
for charge combination $i$ within group $F$.
An event in charge channel $i$ receives weight
$w_F f_{i|F}/q_{i|F}^{\rm Py}$. Since both charge distributions sum to
one, this fixes the division among charge channels while preserving
the group's total probability. The generated momentum distributions
within each charge channel are retained.

An unweighted hadronic decay first selects a separate channel or
fragmentation with probabilities proportional to their partial widths.
Within fragmentation, each trial selects a quark or gluon source with
probability $r_a$ and generates a \textsc{Pythia} event. These events serve
as the proposal sample. We accept an event with probability $w/M$. Here
$w$ is its weight from Eq.~\eqref{eq:weights}, multiplied by the
charge-channel factor where one applies, and $M$ is the largest of these
weights over all allowed groups and charge channels at that mass. A
rejected trial repeats source selection and event generation. Accepted
events follow the target probabilities $P_F$ exactly, and the mean number
of \textsc{Pythia} trials per accepted event is $M$. Reweighting requires a channel that
\textsc{Pythia} can generate; small reference samples give uncertain weights.
The separately treated channels and their
momentum generators are specified in Sec.~\ref{sec:portal-inputs}.
The data-fixed charge fractions used only for the electromagnetic
comparison are specified in Sec.~\ref{sec:em-comparison-definition}.

\paragraph{Separately generated low-multiplicity channels.}
The parameter \texttt{StringFragmentation:stopMass} contributes to the
mass at which \textsc{Pythia} closes a string into its last two hadrons.
Increasing it ends fragmentation earlier and can partially restore
two-body yields while suppressing multihadron production. The size of
this effect depends on mass and channel. We use
\texttt{StringFragmentation:stopMass = 0} for
multihadron generation; constituent endpoint masses still enter the
stopping condition~\cite{Bierlich:2022pfr}.
In the electromagnetic test, $\pi^+\pi^-$, $\pi^0\gamma$,
$\eta\gamma$, $K^+K^-$, $K^0\bar K^0$, $p\bar p$, and $n\bar n$
are calculated separately. Their rates are independent of the
fragmentation parameters used for multihadron production.
Coverage of the separate rates is particle dependent;
Sec.~\ref{sec:portal-inputs} specifies the scalar two-meson extrapolation.

Tab.~\ref{tab:raw-comparisons} collects the definitions and normalizations
of the generator comparisons used throughout this work.

\begin{table*}[!htbp]
\centering
\begin{tabular}{p{0.24\textwidth}p{0.66\textwidth}}
\toprule
Comparison & Generator reference and normalization\\
\midrule
Fig.~\ref{fig:raw-failures}, left panel &
Raw \textsc{Pythia} with electromagnetic charge-squared flavor weights and threshold
factors, multiplied by the inclusive light-quark cross section; channels
forbidden by the current's quantum numbers are retained, and separate
channels are not subtracted.\\
Fig.~\ref{fig:raw-failures}, right panel &
Fractions of raw \textsc{Pythia} events generated from the ALP's $gg$ and
$s\bar s$ sources, combined with their relative probabilities.\\
Figs.~\ref{fig:branchings}, \ref{fig:sm-vector-portals}, and \ref{fig:sm-alp-portal} &
Raw \textsc{Pythia} uses the particle's partonic source weights. At each mass,
our model and raw \textsc{Pythia} share the same total hadronic width;
the raw fractions come directly from the generated channel counts.\\
Fig.~\ref{fig:ship} &
Raw \textsc{Pythia} and our model use the same production,
lifetime, and total hadronic branching ratio. Accepted fractions are
normalized to all hadronic decays inside the decay volume.\\
Fig.~\ref{fig:em-repair} &
The raw reference is the same as in Fig.~\ref{fig:raw-failures}, left panel.
Solid multihadron curves use the tuned probabilities and the width
left after separately calculated channels are subtracted; solid
two-body curves use separate form factors.\\
\bottomrule
\end{tabular}
\caption{Definitions of the \textsc{Pythia} comparisons used with the
event-generation prescription in Sec.~\ref{sec:event-generation}.
Raw \textsc{Pythia} uses the
generator's unmodified fragmentation parameters.}
\label{tab:raw-comparisons}
\end{table*}

\section{Electromagnetic calibration}
\label{sec:currents}

Once each channel group's rate and its division among charge
combinations are fixed at one mass,
does the common prescription reproduce the subsequent mass dependence?
We compare the power law alone with the same law including $T_F$.

\subsection{Calibration and comparison definition}
\label{sec:em-comparison-definition}
The isovector light, isoscalar light, and isoscalar strange components
use fixed weights of approximately $0.75$, $0.08$, and $0.17$,
respectively. These follow from the quark charges: in the isospin
basis, the electromagnetic current has $\rho$-like, $\omega$-like, and
$\phi$-like components with squared coefficients $(e_u-e_d)^2/2$,
$(e_u+e_d)^2/2$, and $e_s^2$, in the ratio $9:1:2$. The weights are
evaluated for the width left after subtracting the separately calculated
channels at the reference mass, which changes them by about one percent.
The first two components are generated with equal $u\bar u$ and
$d\bar d$ samples, and the third with $s\bar s$; the current selection
is applied as described in Sec.~\ref{sec:conditional-model}.

The probability reference mass $\mstar=2.32$~GeV is the lowest tested mass with
normalization inputs and generated events in all ten channel groups,
above the prominent lighter-vector peaks. Lower masses are more sensitive
to thresholds and resonance structure~\cite{BESIII:2022wxz}.
An exploratory comparison using the full \textsc{Pythia} residual $C_F^{\rm Py}$
of Eq.~\eqref{eq:py-residual} gave poorer agreement when the reference mass
was lowered to 2~GeV. The adopted prescription uses $T_F$.

The measurements are from BaBar and
BESIII~\cite{BaBar:2004ytv,BaBar:2006vzy,BaBar:2012sxt,BaBar:2017zmc,
BESIII:2019gjz,BESIII:2020vtu,BESIII:2021ftf,BESIII:2024okl}.
The normalization points at 2.32~GeV are excluded from the comparison
counts in Tab.~\ref{tab:em-scores}.
For this comparison, each channel-group total is split into its measured charge
channels using their fixed relative probabilities at the reference mass.
For example, $3(\pi^+\pi^-)$ and $2(\pi^+\pi^-)2\pi^0$ receive fractions
approximately $0.22$ and $0.78$ of the six-pion total, respectively. The comparison tests the
mass dependence with a data-normalized charge composition.
Without additional charge-channel reweighting, the event generator retains
\textsc{Pythia}'s relative charge probabilities within each group for each
quark-current component. Reproducing the electromagnetic channel split
at event level requires the weight
$P_F f_{i|F}(\mstar^2)/Q_i^{\rm Py}$, where $f_{i|F}$ is the fixed
fraction of charge channel $i$ in group $F$ and $Q_i^{\rm Py}$ is its
generated probability.

\paragraph{Six-pion charge ratio.}
Holding this ratio fixed is supported by BaBar's approximately constant
$\sigma[2(\pi^+\pi^-)2\pi^0]/\sigma[3(\pi^+\pi^-)]=3.98\pm0.06\pm0.41$
outside the structure near 1.6~GeV (Fig.~20 of Ref.~\cite{BaBar:2006vzy}).
There is also a statistical-isospin motivation. With equal pion masses
and equal phase-space integrals of the squared matrix elements, the
identical-particle factors give the inverse ratio
$(2!)^3/(3!)^2=2/9$. Restricting the statistical ensemble to $I=1,I_3=0$
and assigning equal incoherent weights to all independent isospin states
gives $5/18\simeq0.278$. This follows from the projector construction in
Eq.~\eqref{eq:pion-charge-probability}, before excluding measured charge modes.
Isospin fixes the allowed total-isospin sector; equal populations of its
independent states are an additional assumption. We retain the
data-normalized charged-to-neutral ratio $0.2745$ throughout this comparison.
Resonant subchannels can change its mass dependence: BaBar finds that
subtracting the $\omega$-containing contribution removes the approximate
constancy. The fixed charge fractions used here apply to the electromagnetic
calibration; the particle-specific event prescriptions are given in
Sec.~\ref{sec:portal-inputs}.

The comparison covers 14 measured charge channels. We exclude the
BaBar $K^+K^-\pi^0$ dataset with $M(K^+K^-)>1.045$~GeV
\cite{BaBar:2007ceh} from the like-for-like comparison because our
prediction includes the full charge channel without this mass cut.
We do not discard measurements because their quoted uncertainties are large.
The first row of
Tab.~\ref{tab:em-scores} further restricts this selection to energies below the threshold
for producing a pair of charmed mesons and excludes the charmonium
windows at approximately 3.05-3.15 and 3.64-3.73~GeV.
Tab.~\ref{tab:em-scores} also reports the full collection, including the
cut-dependent dataset, resonance regions, and higher energies.
A point is within a factor of two when the prediction
divided by the measured central value lies between $1/2$ and $2$.
These measurements were used to develop the probability prescription.
The comparison shows how closely the model reproduces them.

\begin{figure}[t]
\centering
\includegraphics[width=0.95\textwidth]{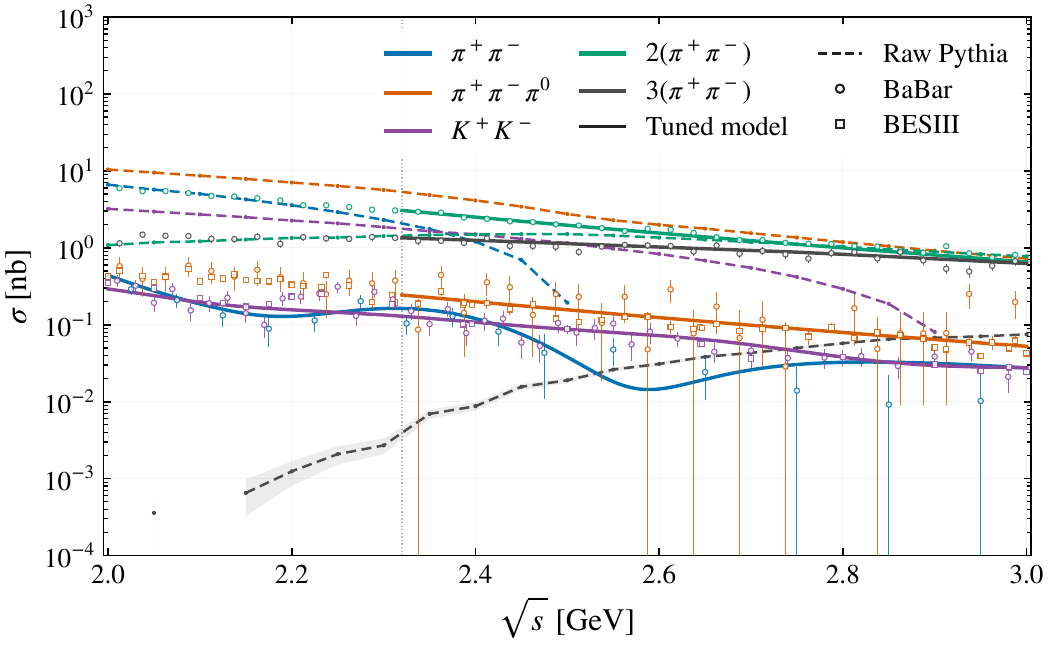}
\caption{Electromagnetic comparison of the model predictions, raw
\textsc{Pythia}, and data for five final states, using the comparison
defined in Sec.~\ref{sec:em-comparison-definition}. Solid and dashed curves
show the model predictions and raw generator; circles and squares show
BaBar~\cite{BaBar:2004ytv,BaBar:2006vzy,BaBar:2012sxt,
BaBar:2012bdw,BaBar:2013jqz} and
BESIII~\cite{BESIII:2018ldc,BESIII:2019gjz,BESIII:2024okl} data.
The three-, four-, and six-pion solid curves begin
at 2.32~GeV. The two-body solid curves use their separate form factors.
Bands show raw-generator Monte Carlo errors. The multihadron model curves
are normalized to data at 2.32~GeV as described above; no additional
rescaling is applied.
\label{fig:em-repair}}
\end{figure}

Tab.~\ref{tab:raw-comparisons} specifies the generator references used
in the figures. Fig.~\ref{fig:em-repair} also shows separate two-body
calculations alongside the tuned three-, four-, and six-pion curves.
The complete set of channel-comparison plots includes every measured channel and the additional
BaBar $K^+K^-\pi^+\pi^-\pi^0$ spectrum~\cite{BaBar:2007qju}.
The complete set is included in the \exhad repository~\cite{exhad}.
The comparator labeled Selected \textsc{Pythia} uses current weights from the widths left after
subtracting separately calculated modes, whereas the tuned and
pure-power curves use the fixed weights stated above. Selected \textsc{Pythia}
rejects channels forbidden by the current's quantum numbers and uses
\texttt{stopMass}=0; it therefore differs from Raw \textsc{Pythia} before the
channel normalization and mass-dependent corrections.

\subsection{Agreement with data}
\label{sec:em-agreement}
The improvement in Tab.~\ref{tab:em-scores} identifies the role of the
finite-mass correction. A pure power fixed at 2.32~GeV falls too rapidly
for the six-pion group. The saturating rise gives the additional
probability required in the data while preserving the chosen high-mass
power. Groups whose generator residual decreases use $T_F=1$.
The correction is therefore selected independently for each group by
the same generator-only rule.

\begin{table*}[t!]
\centering
\begin{tabular}{lrr}
\toprule
Data included & \shortstack{Power law\\alone ($T_F=1$)} &
\shortstack{Power law with\\finite-mass correction}\\
\midrule
Below charm threshold, excluding charmonium & 331/397 & 389/397\\
All energies, excluding $K^+K^-\pi^0$ with a kaon-pair mass cut & 574/764 & 705/764\\
All collected measurements & 618/831 & 749/831\\
Six-pion channels tested separately & 51/98 & 98/98\\
Sum of the two six-pion cross sections & 19/42 & 42/42\\
\bottomrule
\end{tabular}
\caption{Number of data points reproduced within a factor of two / number
tested. Both predictions use Eq.~\eqref{eq:portable-law} with $p=2$ and identical
normalizations and current weights; only the finite-mass factor $T_F$
differs. The first two rows omit the 67 BaBar $K^+K^-\pi^0$ points with
$M(K^+K^-)>1.045$~GeV, a cut not applied to our predictions. The first
also excludes energies above the charm threshold and the charmonium
windows specified in Sec.~\ref{sec:em-comparison-definition}.
The last two rows use this first selection for the two measured six-pion
charge channels: first testing each channel separately, then adding their
cross sections in common energy bins above 2.5~GeV.}
\label{tab:em-scores}
\end{table*}

The largest remaining discrepancy is $K^+K^-\pi^+\pi^-\pi^0$.
Its 37 BESIII points above approximately 3.77~GeV remain included in the complete score.
The measured $s^{-2.49}$ slope corresponds approximately to
$p=1.49$ in Eq.~\eqref{eq:portable-law} after division by an inclusive
cross section proportional to $s^{-1}$~\cite{BESIII:2021ftf}.
This identifies a limitation of the common continuation in this channel.
Its predicted probability is about 1.5\% at the matching mass and below
1\% near 4~GeV, which limits its contribution to inclusive event populations.

\subsection{Cross-section conventions}
\label{sec:em-conventions}
We compare cross sections with photon vacuum-polarization effects removed.
The left panel of Fig.~\ref{fig:raw-failures} and Fig.~\ref{fig:em-repair} use BaBar's
bare $\pi\pi$ and $K\bar K$ cross sections, including final-state
radiation~\cite{BaBar:2012bdw,BaBar:2013jqz}, the undressed
$4\pi$ column~\cite{BaBar:2012sxt}, and BESIII's Born $3\pi$ and
$K\bar K$ cross sections~\cite{BESIII:2019gjz,BESIII:2024okl,BESIII:2018ldc}.
These entries already use the required convention.
The BaBar $3\pi$ and $6\pi$ cross sections in
Refs.~\cite{BaBar:2004ytv,BaBar:2006vzy} include photon vacuum polarization.
We multiply those values and errors by the interpolated ratio
$\sigma_{4\pi}^{\rm undressed}/\sigma_{4\pi}^{\rm dressed}$ from
Tab.~I of Ref.~\cite{BaBar:2012sxt}. This uses its rounded tabulated
values to approximate the common correction $|\alpha(0)/\alpha(s)|^2$.
We apply no further final-state-radiation correction.
\clearpage
\section{Particle-specific inputs and matching}
\label{sec:portal-inputs}

Each particle requires its own exclusive rates, partonic source mixture,
and symmetry restrictions. Tab.~\ref{tab:portals} summarizes the
calculations used for the benchmarks. Tab.~\ref{tab:matching-map} collects their matching intervals.
The mass used to evaluate the fragmentation probabilities is
the parent mass for a boson and the hadronic invariant mass $W$ in an
HNL decay. The transition masses follow the reach of the exclusive
calculations. Meson excitations absent from these calculations remain a source of
uncertainty in the continuation.

\begin{table}[t!]
\centering
\begin{tabular}{@{}p{0.20\textwidth}@{\hspace{1.5em}}p{0.32\textwidth}@{\hspace{1.8em}}p{0.36\textwidth}@{}}
\toprule
{\raggedright Particle or current\par} & {\raggedright Channel-probability and source inputs\par} & {\raggedright Separately calculated channels\par}\\
\midrule
{\raggedright Dark photon\par} & {\raggedright Electromagnetic data and exclusive form factors, with charge-weighted
quark sources~\cite{Ilten:2018crw,Foguel:2022ppx}\par} & {\raggedright Two-body, radiative, nucleon-pair, and $K\bar K\pi$ channels;
charm and nonhadronic decays\par}\\
$B-L$ vector &
{\raggedright Isoscalar exclusive rates at 2~GeV from electromagnetic amplitudes and data
\cite{Ilten:2018crw,Foguel:2022ppx}\par} & {\raggedright $\pi^0\gamma$, $\eta\gamma$, $K\bar K$, nucleon pairs, leptons, and neutrinos\par}\\
Fermion-coupled ALP &
{\raggedright Exclusive pseudoscalar decays at $\approx1.9$~GeV; $gg$ and $s\bar s$ source widths
\cite{Ovchynnikov:2025gpx}\par} & {\raggedright $p\bar p$, $n\bar n$, charm, and nonhadronic decays\par}\\
Higgs-like scalar &
{\raggedright Four-pion composition at 2~GeV; scalar $gg$ and $s\bar s$ widths
\cite{Blackstone:2024ouf,Winkler:2018qyg,Boiarska:2019jym}\par} & {\raggedright $\pi\pi$, $K\bar K$, nucleon pairs, charm, and nonhadronic decays\par}\\
HNL $CC_{ud}$ &
{\raggedright Vector and axial $\tau$ spectra at $W=1.65$~GeV
\cite{ALEPH:2005qgp,Davier:2013sfa}\par} & {\raggedright One-meson decays and other flavor currents
\cite{Bondarenko:2018ptm}\par}\\
HNL $NC_{ud}$ &
{\raggedright Isovector vector and axial spectra, and isoscalar vector form factors,
continued to $W=3$~GeV
\cite{Davier:2013sfa,Ilten:2018crw}\par} & {\raggedright One-meson decays and strange-current contributions
\cite{Bondarenko:2018ptm}\par}\\
\bottomrule
\end{tabular}
\caption{Particle-specific decay inputs described in Sec.~\ref{sec:portal-inputs}.
The separate channels retain their own partial widths and event
generators. $CC$ and $NC$ denote charged and neutral weak currents.}
\label{tab:portals}
\end{table}

\begin{table}[t!]
\centering
\begin{tabular}{p{0.15\textwidth}p{0.18\textwidth}p{0.15\textwidth}p{0.39\textwidth}}
\toprule
Particle & Quantity & Interval [GeV] & Connection\\
\midrule
ALP & $P_F$, including $P_R$ & $\approx1.9$--3 &
Exclusive probabilities to the power-law continuation; values and slopes\\
$B-L$ & $\Gamma_{\rm had}$ & 1.70-1.78 &
Short smooth connection at the exclusive-sum/perturbative-width crossing\\
$B-L$ & $\Gamma_F$ & 2-3 &
Exclusive partial widths to the power-law continuation;
the total width is fixed by perturbative QCD\\
Dark photon & $\Gamma_{uds}$ & 1.70-2 &
Exclusive sum to perturbative inclusive width; smooth interpolation\\
Dark photon & Seven $\Gamma_F/\Gamma_{uds}$ & 2-2.5 &
Exclusive probabilities to electromagnetic fits; values and slopes\\
Dark photon & Events in $R$ & Starts at 1.70 &
\textsc{Pythia} generates decays into channels outside the known exclusive calculation\\
Scalar & $P_F$ & Starts at 2 &
Boundary values followed directly by the power law and finite-mass factors;
no separate slope-matching interval\\
\bottomrule
\end{tabular}
\caption{Matching prescriptions described in Sec.~\ref{sec:portal-inputs}.
The channel probabilities $P_F=\Gamma_F/\Gamma_{\rm frag}$ are defined in
Eq.~\eqref{eq:conditional-definition}. Each interpolation
uses the quantities calculated at its endpoints.}
\label{tab:matching-map}
\end{table}

\subsection{Dark photon}
\label{sec:dark-photon-inputs}

The exclusive decay inputs follow Ref.~\cite{Ilten:2018crw}
and DeLiVeR~\cite{Foguel:2022ppx}.
A dark photon couples to the electromagnetic current, so the measured
exclusive cross section fixes the corresponding partial width:
\begin{equation}
 \frac{\Gamma(A'\to F)}{\Gamma(A'\to\mu^+\mu^-)}
 =\frac{\sigma(e^+e^-\to F)}{\sigma(e^+e^-\to\mu^+\mu^-)}
 \quad\text{at }s=m_{A'}^2 .
 \label{eq:dark-photon-data}
\end{equation}
The cross sections have photon vacuum-polarization effects removed.
Well-measured
exclusive states account for a large, mass-dependent fraction of the
hadronic width. We use these measurements directly for their channel
probabilities. The electromagnetic comparison in
Sec.~\ref{sec:currents} tests whether the portable prescription can
describe the same spectra with a restricted set of assumptions.
At higher masses, the measured exclusive contributions fall and an
increasing fraction of the inclusive width remains in channels without
exclusive data. Their probabilities and momenta still require a model.
Our model therefore combines measured rates, exclusive form-factor
calculations, and generated events for unmeasured channels.

Let $\Gamma_{uds}$ denote the inclusive light-hadronic width, excluding charm.
Below 1.70~GeV, exclusive form factors determine the partial widths.
Between 1.70 and 2~GeV, the exclusive calculation fixes the channel
fractions $\Gamma_F/\Gamma_{uds}$ while the total light-quark width approaches its
perturbative value smoothly. The high-mass description uses seven fits
to BaBar and BESIII data, constrained to give nonnegative cross sections,
for $3\pi$,
the all-charged and neutral-pion-containing
$4\pi$ and $6\pi$ states, and the all-charged and neutral-particle-containing
$K\bar K\pi\pi$ states. Resonance and smooth amplitudes are combined
before squaring, with fixed pole masses and widths.
An ordinary cubic interpolation joins the exclusive fractions $\Gamma_F/\Gamma_{uds}$ and
their derivatives at 2~GeV to the fit fractions and derivatives at
2.5~GeV. Above 2.5~GeV, these rates follow the fits directly.
The two-body, radiative, nucleon-pair, and $K\bar K\pi$ channels
retain their separate form-factor continuations.

The 1.70-2~GeV interpolation concerns the inclusive $u,d,s$ width.
Its lower endpoint follows the intersection of the exclusive and
perturbative light-quark estimates in Ref.~\cite{Foguel:2022ppx}.
It is a modeling choice for the transition between these descriptions.
Charmed-meson pairs become kinematically accessible at
$m=2m_{D^0}\simeq3.73$~GeV; the separate charm contribution is zero
below this threshold. Narrow charmonium resonances, including
$J/\psi$ and $\psi(2S)$ below this threshold, require their resonance
rates and are excluded from the smooth electromagnetic comparison.

The remaining light-hadronic width is fixed by subtraction:
\begin{equation}
 \Gamma_R^{uds}=\Gamma_{uds}-\Gamma_{\rm sep}^{uds}
                         -\sum_{F=1}^{7}\Gamma_F ,
 \label{eq:em-remainder}
\end{equation}
where $\Gamma_{\rm sep}^{uds}$ includes only the separately generated
light-hadronic channels, and $\Gamma_F$ follows the interpolation or the
electromagnetic fit at the relevant mass. We require this difference to
be nonnegative. Charm is treated separately above the threshold for
producing charmed hadrons; its events exclude the measured light final
states to prevent double counting.

After a channel is selected, its primary hadronic state and subsequent
decays determine the momenta at the dark-photon mass. A primary state
specifies the hadrons before their subsequent decays. For example,
$K^{*0}K^-\pi^+$ and its charge conjugate, followed by $K^*\to K\pi$,
and $\phi\pi^+\pi^-$ followed by $\phi\to K^+K^-$ contribute to the
same $K^+K^-\pi^+\pi^-$ channel. Above 2~GeV, their relative
probabilities are held at their 2~GeV values, as for most primary-state
mixtures. Kaon-pair and $K\bar K\pi$ states retain their
mass-dependent form-factor probabilities.
From 1.70~GeV, we generate \textsc{Pythia} events for each quark-current component
and reject states that violate its symmetry restrictions. These events
describe the unmeasured contribution; separately generated channels are
excluded to avoid double counting.

\paragraph{Charge probabilities in the pion remainder.}
Unmeasured pion channels need a charged-versus-neutral composition before
their detector signatures can be predicted. For all-pion states without
an explicitly modeled resonance decay chain, we assign charge probabilities using the pion multiplicity
$n$ and the initiating current's isospin $(I,I_3)$. Without additional
partial-wave or permutation-symmetry information, we assume equal incoherent
weights for all independent isospin-coupling states. The probability
of charge counts $c=(n_+,n_0,n_-)$ is
\begin{equation}
 P_{\rm iso}(c\mid n,I,I_3)=
 \frac{\operatorname{Tr}(\Pi_c\Pi_{I,I_3})}
 {\displaystyle\sum_{c'\in\mathcal A}
  \operatorname{Tr}(\Pi_{c'}\Pi_{I,I_3})},\qquad c\in\mathcal A .
 \label{eq:pion-charge-probability}
\end{equation}
Here $\Pi_c$ selects all charge orderings with counts $c$,
$\Pi_{I,I_3}$ selects the source isospin subspace, and $\mathcal A$
contains kinematically open charge configurations without separately
assigned or fitted rates. This changes the charge fractions while
preserving each source's rate at fixed pion multiplicity. Resonance
decay chains and measured channels retain their probabilities.
For each selected charge configuration, we generate \textsc{Pythia}
events until that configuration is obtained. If the retry limit is
exhausted, pion momenta are sampled in flat $n$-body phase space before
daughter decays; Sec.~\ref{sec:statistical-estimators} specifies the limit.

Below 4~GeV, all eligible pion-remainder events use the statistical-isospin
charge probabilities. Between 4 and 5~GeV, we interpolate between these
probabilities and \textsc{Pythia}'s charge probabilities. This interval
is chosen as part of the model. The statistical-isospin
weight is $1-3x^2+2x^3$, with
$x=(m-4\,\mathrm{GeV})/(1\,\mathrm{GeV})$; the \textsc{Pythia} weight is its
complement. Above 5~GeV, \textsc{Pythia} determines the charge probabilities.
The \textsc{Pythia} component retains the quark-current symmetry restrictions
and excludes separately generated channels. The remainder width and measured-channel rates
are unchanged throughout this transition.
We use this description for the dark-photon SHiP result.

\subsection{\texorpdfstring{$B-L$}{B-L} vector}
\label{sec:hadronic-composition}

We first determine the total hadronic width and the explicitly
calculated channel widths, then assign their difference to the remainder.
The channel rates and total width have distinct matching intervals.

\paragraph{Exclusive inputs.}
With zero kinetic mixing, the light-quark current is
$J_B^\mu=(\bar u\gamma^\mu u+\bar d\gamma^\mu d+
\bar s\gamma^\mu s)/3$.
The exclusive amplitudes follow the flavor decomposition of the
electromagnetic form factors in Refs.~\cite{Ilten:2018crw,Foguel:2022ppx}.
After factoring out the overall gauge couplings, the $\rho$-like amplitude
vanishes. Relative to their electromagnetic values, the $\omega$-like
amplitude is multiplied by two and the $\phi$-like amplitude changes sign.
We combine these amplitudes, including interference, before
calculating the width. Electromagnetic measurements constrain their
sum; the transfer to $B-L$ depends on this flavor decomposition.
For fragmentation, the light and strange isoscalar event samples
have probabilities $2/3$ and $1/3$.

The exclusive calculation covers $3\pi$, $K\bar K\pi$,
$\omega\pi\pi$, $\eta\omega$, $\eta\phi$, $\phi\pi\pi$, and
the $K^*\bar K\pi$ contribution to $K\bar K\pi\pi$
\cite{Foguel:2022ppx}. Resonance decays assign these rates to the
groups $3\pi$, $K\bar K\pi$, $5\pi$, $\eta3\pi$,
$\eta K\bar K$, and $K\bar K\pi\pi$.
In particular, $\omega\pi\pi$ followed by $\omega\to3\pi$ gives
most of the five-pion rate below 2~GeV. Each resonance contribution is
counted once using the channel decomposition of Ref.~\cite{Foguel:2022ppx};
the remaining $\omega$ and $\phi$ daughter modes contribute to $R$.

The $\pi^0\gamma$, $\eta\gamma$, $K\bar K$, $p\bar p$, and
$n\bar n$ rates are calculated separately
\cite{Ilten:2018crw,Foguel:2022ppx,Plehn:2019jeo}.
The radiative modes keep their vector-meson form factors throughout the
displayed range. The charged and neutral kaon widths use the same coherent
form-factor fit of Ref.~\cite{Plehn:2019jeo} throughout this range, summed
without fragmentation reweighting or a change of calculation at the
matching masses. The charged and neutral kaon-pair event probabilities are determined by
these same $B-L$ partial widths.
The nucleon electric and magnetic form factors use the isoscalar
calculation of Ref.~\cite{Plehn:2019jeo}: the $B-L$ amplitude is twice
the isoscalar electromagnetic amplitude in its convention, and the
strange nucleon form factor is neglected.

Near 2~GeV, the $3\pi$ normalization uses a local fit to BaBar and
BESIII measurements~\cite{BABAR:2021cde,BESIII:2024okl}.
A common factor, equal to the fitted electromagnetic cross section
divided by the form-factor prediction, rescales the complete $B-L$
three-pion width. This retains the relative light- and strange-quark
amplitudes and their interference. The local fit changes the
normalization and slope. It holds the relative flavor amplitudes fixed,
so their uncertainties are outside this fit. Tab.~\ref{tab:bl-boundary-inputs} gives the numerical
fit choices. A smooth interpolation over 1.8-1.9~GeV connects this
rate to the exclusive calculation at lower mass.
For $K\bar K\pi$, we use the isoscalar normalization from the BaBar
Dalitz analysis, including its charge sum~\cite{BaBar:2007ceh}, with the
$\phi$-like flavor assignment of Ref.~\cite{Foguel:2022ppx}.
Its experimental uncertainty defines an alternative
normalization at 2~GeV.

\begin{table*}[t!]
\centering
\begin{tabular}{p{0.35\textwidth}p{0.54\textwidth}}
\toprule
Quantity or choice & Value\\
\midrule
Data and convention & BaBar ISR and BESIII scan; Born cross sections,
averaged over each experimental bin\\
Fit interval & 1.8-2.2~GeV\\
Fit function & Quadratic in $\log(m/2\,\mathrm{GeV})$ for $\log\sigma_{3\pi}$\\
Fit quality & $\chi^2/\mathrm{dof}=63.3/20$; covariance multiplied by this ratio\\
$\Gamma_{3\pi}(2\,\mathrm{GeV})/g_{B-L}^2$ &
$3.54\times10^{-3}$~GeV; $(3.31$-$3.68)\times10^{-3}$~GeV
under changes of fit interval, order, and correlated errors\\
\bottomrule
\end{tabular}
\caption{Local three-pion input for the $B-L$ vector, using the fit
described in Sec.~\ref{sec:hadronic-composition}. The uncertainty
range describes alternative fit choices; flavor decomposition adds a
separate model uncertainty. Widths use unit $B-L$ gauge coupling.}
\label{tab:bl-boundary-inputs}
\end{table*}

\paragraph{Matching.}
Let $S(m)$ be the sum of the exclusive widths just described, including
the separate radiative, two-kaon, and nucleon-pair modes. The inclusive
light-quark width is calculated independently as
\begin{equation}
 I(m)=\frac{g_{B-L}^2m}{12\pi}
 \left[1+\delta_{\rm QCD}^{B-L}(m)\right].
 \label{eq:baryon-inclusive}
\end{equation}
Here, $g_{B-L}$ is the gauge coupling, $N_c=3$ counts quark colors,
and $N_c\sum_{q=u,d,s}(1/3)^2=1$. The factor
$\delta_{\rm QCD}^{B-L}$ includes the charge-dependent corrections through
$\alpha_s^4$ specified in Ref.~\cite{Foguel:2022ppx}.
We neglect light-quark masses. Following the hadron-quark transition
criterion of Ref.~\cite{Foguel:2022ppx}, we use the last crossing of $S$
and $I$, near 1.75~GeV for these inputs. We interpolate smoothly between
$S$ and $I$ over 1.70-1.78~GeV using a fifth-degree Hermite polynomial
for $\log\Gamma_{\rm had}$,
matching the values, slopes, and curvatures of $S$ and $I$ at the two
ends. Below this interval, $\Gamma_{\rm had}=S$;
above it, $\Gamma_{\rm had}=I$. The interpolation stays above $S$, so
all calculated exclusive widths fit within the inclusive total.
The multihadron partial widths follow the exclusive calculation through
2~GeV and approach the probability law over 2-3~GeV. The exclusive inputs
continue into this interval: the local three-pion fit extends to 2.2~GeV, and the
form-factor calculations are evaluated through 3~GeV. The upper endpoint
coincides with the start of the 3-5~GeV interval used to extract \textsc{Pythia}'s
high-mass power in Sec.~\ref{sec:finite-mass}. These fragmentation
endpoints are modelling choices; the inclusive width remains fixed
by Eq.~\eqref{eq:baryon-inclusive} throughout this interval.

Let $E_F(m)$ denote the exclusive width of group $F$ evaluated for the $B-L$ current,
continued with the form factors to $m_1=3$~GeV. The three-pion calculation
is joined to its normalized form-factor tail on 2-2.2~GeV; the
$K\bar K\pi$ input is joined on 1.9-2~GeV. Both tails retain their
normalization at 2~GeV. The other multihadron widths are calculated from
the resonance form factors fitted to electromagnetic data in
Refs.~\cite{Plehn:2019jeo,Foguel:2022ppx}, evaluated for the $B-L$ current.
Resonance decays are counted once. The widths at 3~GeV are form-factor
extrapolations. Their uncertainties affect the predicted composition
but are not included in the central curves (Sec.~\ref{sec:uncertainties}).

Write
$G(m)=I(m)-\Gamma_{\rm sep}(m)$ for the width available to
fragmentation after subtracting the radiative, two-kaon and nucleon-pair
modes. For the six specified fragmentation groups, the high-mass
partial-width continuation is
\begin{equation}
 H_F(m)=E_F(m_1)\frac{G(m)}{G(m_1)}
 \left(\frac{m_1}{m}\right)^{2p}
 \frac{T_F(m^2)}{T_F(m_1^2)},\qquad p=2.
 \label{eq:baryon-high-width}
\end{equation}
Dividing by $G(m)$ gives the probability law in
Eq.~\eqref{eq:portable-law}, normalized at $m_1$.
On $2<m<3$~GeV, set $x=(m-2\,\mathrm{GeV})/(1\,\mathrm{GeV})$
and $h(x)=6x^5-15x^4+10x^3$. We use
\begin{equation}
 \Gamma_F=(1-h)E_F+hH_F,\qquad \Gamma_{\rm had}=I.
 \label{eq:baryon-common-width}
\end{equation}
All six groups use the same interpolation, with $E_F$ and $H_F$ evaluated
at the particle mass $m$. Below 2~GeV, $h=0$; above 3~GeV, $h=1$.
Its first and second derivatives vanish at both endpoints. The input
thresholds and resonances are retained. The separate form-factor modes are
unchanged. The remainder is
$\Gamma_{\rm had}-\Gamma_{\rm sep}-\sum_F\Gamma_F$;
we verify that it is nonnegative and that all channel probabilities sum
to one throughout the mass range and for each power variation.

\paragraph{The unmeasured contribution.}
The inclusive light-quark width substantially exceeds the sum of the
calculated exclusive hadronic widths, producing a large Other fraction
at higher masses.
The form-factor calculation of
Ref.~\cite{Foguel:2022ppx} does not determine the composition
of this difference, and the exclusive inventory in
Ref.~\cite{Ilten:2018crw} does not include the
$\omega\pi\pi$ channels needed to check the five-pion rate.
The unmeasured width is already nonzero below 2~GeV. We include it in the
channel-fraction plots, but have no complete event-generation prescription
for this contribution below 2~GeV.

At masses of at least 2~GeV, \textsc{Pythia}
distributes the remainder among many multihadron states; its largest
individual charge channel accounts for about six percent of the
remainder at 3~GeV.
The total of this contribution follows from subtraction, while its
composition is modeled. The remainder also includes smaller contributions
from decays of the $\omega$ and $\phi$ in $\omega\pi\pi$, $\eta\omega$,
$\phi\pi\pi$, and $\eta\phi$, such as
$\omega\pi\pi\to\pi^0\gamma\pi\pi$.
Their contributions to the mediator's final-state probabilities are joined
over 2--3~GeV to the common law in Eq.~\eqref{eq:portable-law} with $T_F=1$;
the $\omega$ and $\phi$ decay branching fractions remain fixed.

The full $B-L$ width also includes charged leptons and three light
neutrinos. Each neutrino species contributes half the width of a
massless charged Dirac fermion with unit charge
\cite{Ilten:2018crw}. The benchmark assumes no additional accessible
particles. Above the threshold for producing charmed hadrons, the charm width must enter
separately; the figures here remain below that threshold.

\subsection{Fermion-coupled ALP}
\label{sec:alp-inputs}

We consider an ALP with universal fermion couplings specified at 1~TeV.
Our decay inputs use the calculation of Ref.~\cite{Ovchynnikov:2025gpx};
see also Refs.~\cite{Aloni:2018vki,Balkin:2025enj}.
The adopted calculation includes pseudoscalar-meson mixing, the
dominant $\eta^{(\prime)}\pi\pi$ and $K\bar K\pi$ modes, and the
mass-dependent gluon and strange-quark widths. The latter determine the
relative $gg$ and $s\bar s$ event samples. The exclusive probabilities
and their boundary slopes determine the left endpoint of
Eq.~\eqref{eq:alp-probability-match}; Tab.~\ref{tab:numerical-boundaries}
specifies the numerical continuation to that endpoint.

We interpolate the channel-group probabilities $P_F$ and the remainder
probability $P_R$ together from $\mstar\approx1.9$~GeV to 3~GeV, preserving their
normalization, nonnegativity, and endpoint values and slopes.
Sec.~\ref{sec:interpolation-details} gives the coefficient construction
in Eq.~\eqref{eq:alp-probability-match}.
The fifth-degree interpolation fixes the shape within this interval
as a modelling assumption. Above 3~GeV, the channel-group probabilities
$P_F$ follow Eq.~\eqref{eq:portable-law}.

The charge probabilities $f_{i|F}$ are matched separately from the
group totals. For an isoscalar ALP, the dipion in
$\eta^{(\prime)}\pi\pi$ has $I=0$. In the equal-pion-mass limit,
\begin{equation}
 \frac{\Gamma(a\to\eta^{(\prime)}\pi^+\pi^-)}
      {\Gamma(a\to\eta^{(\prime)}\pi^0\pi^0)}=2.
 \label{eq:alp-isoscalar-dipion}
\end{equation}
We impose this ratio separately for $\eta\pi\pi$ and $\eta'\pi\pi$
at and above 3~GeV, preserving the total probability of each mode.
Within the existing $\mstar$--3~GeV interval, the fifth-degree
matching retains the exclusive charge probabilities and their slopes
at $\mstar$, including the small charged--neutral pion-mass effects,
and ends at the isospin-constrained probabilities and slopes.
This continuation of the finite-mass corrections is a modelling
assumption; above 3~GeV the charge ratio uses the equal-mass approximation.
The $K\bar K\pi$ charge probabilities retain their matching to the
\textsc{Pythia} values and slopes at 3~GeV.

The proposal probabilities $q_{i|F}^{\rm Py}$ use the source mixture
of Eq.~\eqref{eq:source-mixture} and normalized monotone cubic
interpolation between simulated masses. The generator applies
$f_{i|F}/q_{i|F}^{\rm Py}$ as described in
Sec.~\ref{sec:event-generation}; these weights remain active above
3~GeV for $\eta^{(\prime)}\pi\pi$. From $\mstar$, \textsc{Pythia}
generates the momenta within each charge channel. The charge correction
leaves all channel-group probabilities and total widths unchanged.

At the matching mass, the specified groups exhaust $\Gamma_{\rm frag}$,
so $P_R=0$. Their decreasing probabilities leave room for additional
hadronic states at larger masses.

The nucleon-pair widths remain separately calculated across this
transition, with their own narrow interpolation. The total and nonhadronic widths retain the
mass dependence of Ref.~\cite{Ovchynnikov:2025gpx}.
Charm is generated separately above its physical threshold.

The small isospin-breaking $3\pi$ contribution and the
$\pi^+\pi^-\gamma$ contribution each have their own probability in
Eq.~\eqref{eq:alp-probability-match} and a separate momentum generator.
For $3\pi$, the relative probabilities of $\pi^+\pi^-\pi^0$ and $3\pi^0$
are fixed at the exclusive boundary, and a constant three-body matrix element determines
the momenta. For $\pi^+\pi^-\gamma$, the generator retains the anomalous
factor $|\mathbf p_+\times\mathbf p_-|^2$, evaluated in the parent rest
frame. Beyond this explicit momentum factor, the remaining form factor
is held constant, neglecting its dependence on the decay invariant masses.
These choices approximate the distributions above the exclusive region. Additional particles,
as in $3\pi\gamma$ or $K\bar K\eta$, place an event in a different group
or in the remainder.

\subsection{Higgs-like scalar}
\label{sec:scalar-inputs}

The scalar form factors of Ref.~\cite{Blackstone:2024ouf}
determine the two-pion and two-kaon rates. We use the central, lower, and upper
prescriptions of that calculation, together with the alternative of
Ref.~\cite{Winkler:2018qyg} and the corrections and additional channels
of Ref.~\cite{Boiarska:2019jym}. These calculations also specify the
total width and the gluon and strange-quark contributions used for
fragmentation. Their spread describes uncertainty in the scalar rates,
separately from variation of the common power $p$.

After the separate two-meson and nucleon-pair rates are subtracted,
the exclusive input represents the remaining light-hadronic width
by four-pion modes below 2~GeV. We adopt 2~GeV as a phenomenological
boundary between the exclusive and fragmentation descriptions.
This choice sets $b_{4\pi}=1$ and
$b_R=0$ at the matching point.
The power law and finite-mass factor apply directly above this mass.
The scalar probabilities follow Eq.~\eqref{eq:portable-law} immediately,
without a separate interval for matching their slopes.

The two-meson rate tables end immediately below 2~GeV, at the endpoint
$m_{\rm end}$ specified in Tab.~\ref{tab:numerical-boundaries}. To retain these
open channels above that mass, we hold each two-body amplitude at its
endpoint value and include the changing two-body phase space. For a pion
or kaon $h$ of mass $m_h$ in a specified charge channel, the continued width is
\begin{equation}
 \Gamma_{hh}(m)=\Gamma_{hh}(m_{\rm end})\frac{m_{\rm end}}{m}
 \frac{\beta_h(m)}{\beta_h(m_{\rm end})},\qquad
 \beta_h(m)=\sqrt{1-\frac{4m_h^2}{m^2}} .
 \label{eq:scalar-two-meson-tail}
\end{equation}
This extrapolation holds the endpoint amplitudes fixed through 5~GeV.
It is applied separately to each pion and
kaon charge channel and each of the four rate prescriptions. For each
prescription, the total hadronic, leptonic, nucleon-pair, and charm widths
retain their original mass dependence. Only the two-meson amplitudes
receive the continuation defined above.
Subtracting the continued two-meson widths leaves the width assigned
to fragmentation. The short numerical connection in
Tab.~\ref{tab:numerical-boundaries} assigns this remainder to four pions,
preserving the exclusive calculation's relative probabilities for the
four-pion charge combinations. At
2~GeV, it therefore joins continuously to $P_{4\pi}=1$ in the
fragmentation model. The figure and event generator use the same rates.

\subsection{Heavy neutral leptons}
\label{sec:hnl-inputs}

The accompanying lepton leaves a variable invariant mass $W$ for the
hadronic system. We therefore construct hadronic probabilities at fixed
$W$ and integrate them with the weak decay spectrum at each parent mass.
The calculation of Ref.~\cite{Bondarenko:2018ptm} determines the weak decay
normalizations, mixing dependence, and one-meson rates.
Ref.~\cite{Feng:2024zfe} also retains one-meson modes and estimates the
multihadron remainder by subtracting their summed width from the
inclusive quark-level width.

We match exclusive and inclusive widths independently
for each charged- or neutral-current quark-flavor sector and each mixing
component $\alpha=e,\mu,\tau$. For sectors with exclusive inputs, the
matching mass is the parent mass $m_N$ where the inclusive quark-level
width overtakes the summed exclusive hadronic width. Below this crossing,
the exclusive description is used. Above it, the separately generated
meson--lepton or meson--neutrino modes are retained, and their summed width
is subtracted from the inclusive width of that same sector and mixing
component. Only the remainder is assigned to multihadron generation;
the low-mass multihadron rates are replaced, not added again. The tables
for the three mixing components are then combined with weights
$|U_\alpha|^2$. There is no universal HNL switching mass: for example,
the light neutral-current crossing is at $m_N\simeq1.380$~GeV,
whereas the $u\bar d$ charged-current crossings are approximately
$1.848$, $1.865$, and $3.628$~GeV for electron, muon, and tau mixing.
These parent-mass crossings fix the input widths; they are distinct from
the boundaries in $W$ used below to construct the multihadron composition.
\textsc{exHad} changes that composition without changing the input widths.

The nonstrange
charged-current multihadron probabilities follow the vector and axial
spectral functions measured by ALEPH in nonstrange hadronic $\tau$
decays~\cite{ALEPH:2005qgp}, with the updated spectral functions of
Ref.~\cite{Davier:2013sfa}. These spectral functions encode the distribution
in hadronic invariant mass for each current. For the neutral current, isospin
relates its isovector vector and axial parts to the $\tau$ spectra,
with the corresponding weak coefficients. The isoscalar vector part
uses electromagnetic form factors from \textsc{DarkCast}
\cite{Ilten:2018crw}.
For the charged-current mass distribution, the sum of the vector and
axial spectral functions is held at its $W=1.65$~GeV value above that
mass. For the neutral current, the isovector vector and axial spectral
functions are linearly interpolated from their 1.65~GeV values to a
common constant at 2~GeV, then held fixed. This constant is the average
of their two median values over $1.3\leq W\leq1.6$~GeV. These are
assumptions for extrapolation beyond the measured range.

Here $CC$ and $NC$ denote charged and neutral weak currents, respectively;
the subscript $ud$ identifies the nonstrange $u,d$ sector.
The $CC_{ud}$ and $NC_{ud}$ probabilities are normalized at the hadronic
masses in Tab.~\ref{tab:generator-reference-masses}, with the narrow
charged-current connection specified in Tab.~\ref{tab:numerical-boundaries}.
These probability continuations are evaluated through $W=5$~GeV.
Sec.~\ref{sec:hnl-numerical-range} specifies the larger mass range supported
by the HNL event generator. The $CC_{us}$ contribution retains its
$K\pi$ and $K\pi\pi$ resonance description constrained by $\tau$
decays~\cite{ALEPH:2005qgp,OPAL:1998rrm}. Strange neutral currents and
the charged-charm currents $CC_{cd}$ and $CC_{cs}$ have separate
rates and flavor-specific event generation.

The HNL panels integrate the channel probabilities over $W$ with the
differential weak decay rate at each parent mass $m_N$. They describe
the indicated nonstrange multihadron current. The complete HNL decay
distribution also includes one-meson and other flavor channels.
The use of isospin, continuation above the measured $\tau$ range, and
the experimental spectral uncertainties limit the accuracy of these
predicted fractions.
The fragmentation variations hold the spectral functions fixed and
therefore do not propagate their experimental uncertainties or correlations.

\paragraph{Comparison with resonance-resolved HNL calculations.}
Ref.~\cite{Schubert:2026mgf} provides meson-pair and three-pion
amplitudes and resonant four-body event samples (App.~B).
Its light-current multihadron prescription uses hadronic amplitudes below
$q_H=1.25$~GeV and leading-order quark widths above (Sec.~5).
This boundary in hadronic invariant mass differs from our sector- and
mixing-dependent parent-mass crossings. Its illustrative continuation uses
untuned \textsc{Pythia} fractions that disagree with the exclusive
calculation (App.~C).

Our prescription continues the current-specific channel
probabilities with Eq.~\eqref{eq:portable-law} at $s=W^2$, using the
common exponent tested against electromagnetic data. We determine the
normalization, charge fractions, and finite-mass factors for each weak
current. \textsc{EventCalc-SHiP} retains its tabulated
current widths and lifetime, so the HNL panels test the decay composition
with these inputs held fixed.
Our reweighting preserves the generated
kinematics within each channel group. A complete treatment of resonance
interference and hadronic polarization correlations requires an
amplitude-level description.

\section{Decay composition and search signatures}
\label{sec:decay-results}

\subsection{Channel fractions and their physical origin}
\label{sec:portal-branching-plots}

\begin{figure}[!htb]
\centering
\resizebox{\textwidth}{!}{%
\includegraphics{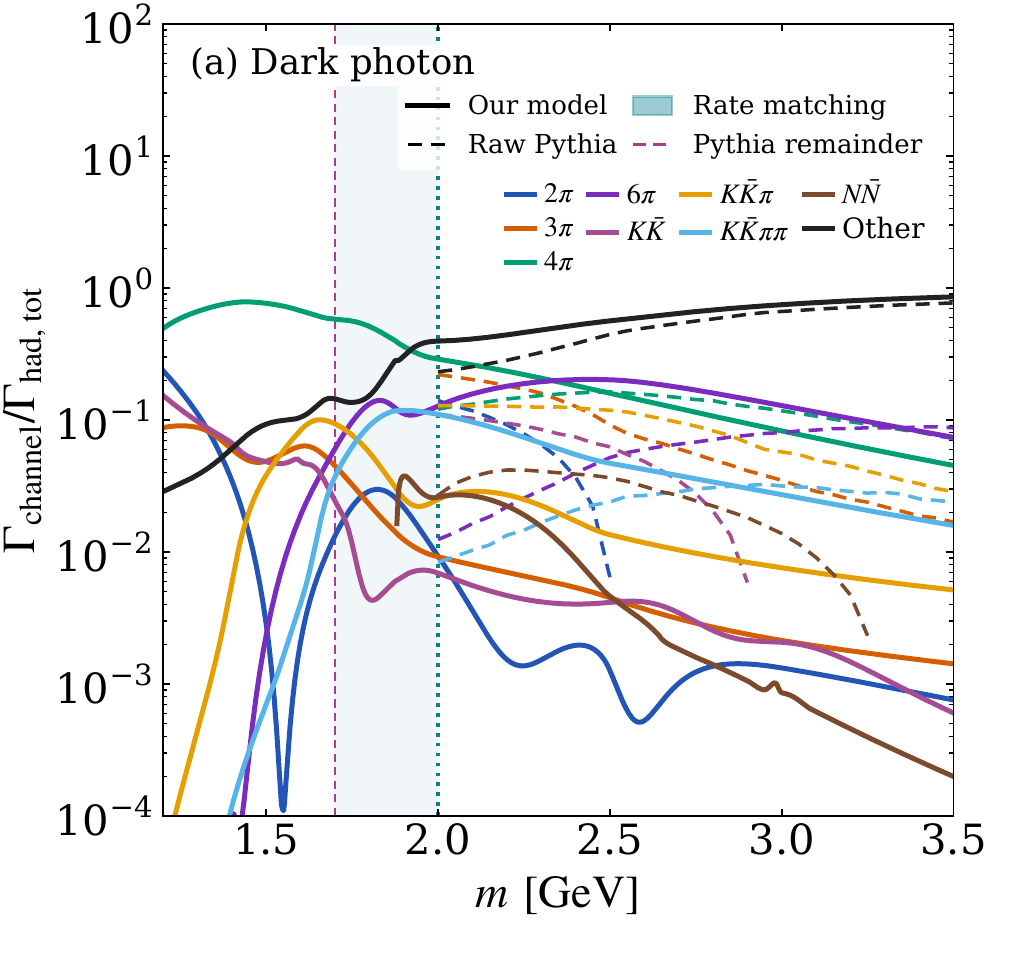}%
\includegraphics{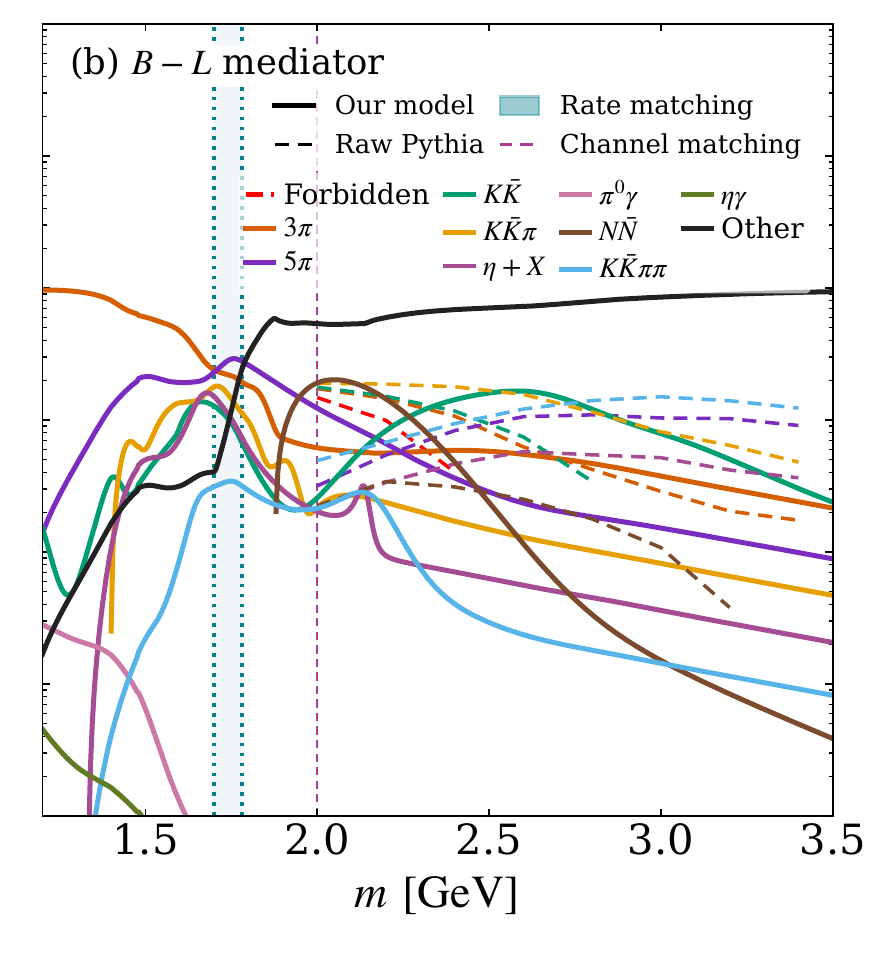}%
}
\caption{Hadronic channel fractions for a dark photon (left) and a
$B-L$ vector (right), as defined in Sec.~\ref{sec:portal-branching-plots}.
Solid curves show our model; dashed curves show raw \textsc{Pythia}. The curves show
fractions of all hadronic decays, including radiative hadronic modes.
Above the 1.70--1.78~GeV matching interval, the total $B-L$ hadronic width
in the denominator is given by Eq.~\eqref{eq:baryon-inclusive}.
The remainder is obtained by subtracting the separately generated and six
specified multihadron channel widths from this total. It includes the
smaller $\omega$ and $\phi$ decay contributions described in
Sec.~\ref{sec:hadronic-composition}.
Separate raw-\textsc{Pythia} curves are not shown for the $B-L$ $\eta\gamma$
contribution or the subdivisions of its remainder; these appear only
as solid curves.
The displayed masses lie below the threshold for producing a pair of charmed mesons.
\label{fig:sm-vector-portals}}
\end{figure}

\begin{figure}[!htb]
\centering
\includegraphics[width=0.80\textwidth]{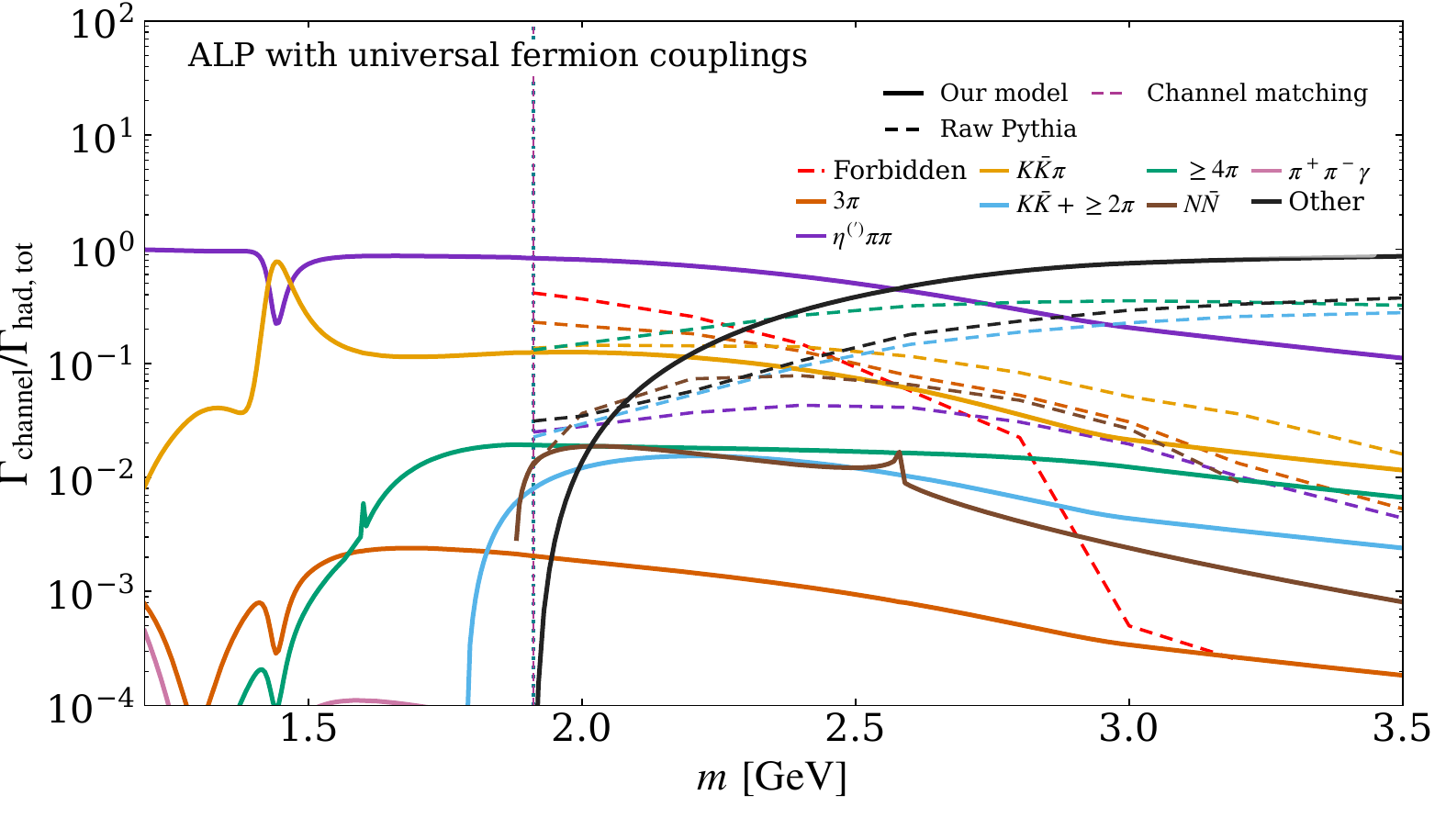}
\caption{Hadronic channel fractions for an ALP with universal fermion
couplings at 1~TeV. Solid curves combine the exclusive calculation
of Ref.~\cite{Ovchynnikov:2025gpx} with the tuned continuation;
dashed curves show raw \textsc{Pythia}. Nucleon pairs retain their
separate calculated widths. The Forbidden curve is present only for raw
\textsc{Pythia} and contains two-pseudoscalar states, including $\pi\pi$,
$K\bar K$, and $\eta\eta$. The denominator is the total hadronic width,
including hadrons with photons. The displayed masses lie below the
threshold for producing a pair of charmed mesons. The matching prescription is given in
Sec.~\ref{sec:alp-inputs}.
\label{fig:sm-alp-portal}}
\end{figure}

\begin{figure}[!htb]
\centering
\includegraphics[width=0.80\textwidth]{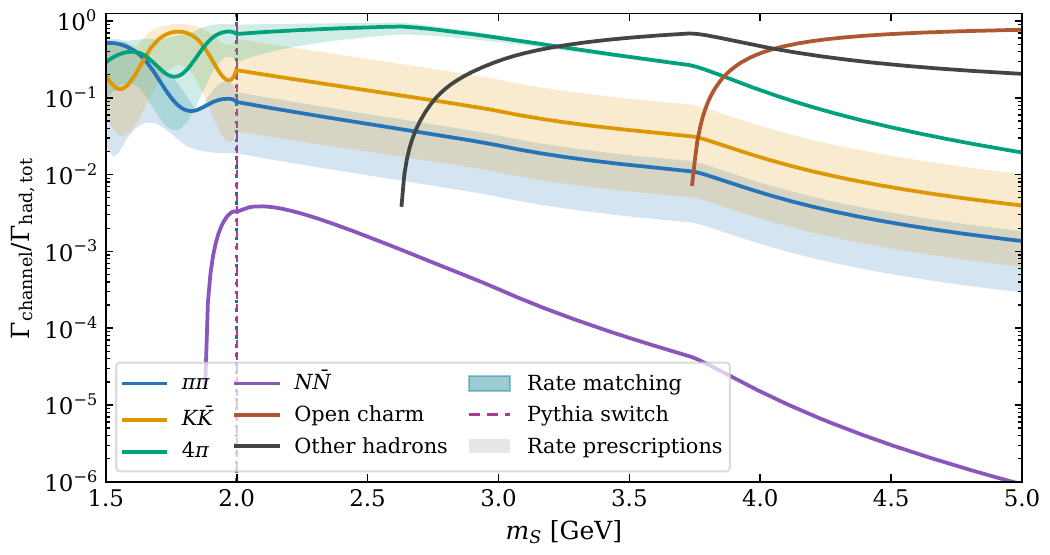}
\caption{Hadronic channel fractions for a Higgs-like scalar, obtained
from the central exclusive rate calculation and the tuned
fragmentation probabilities. The pion-pair, kaon-pair, and nucleon-pair
rates are calculated separately; $\Gamma_{\rm frag}$ is distributed
between the four-pion group and the remaining multihadron channels.
The charm contribution is separate. Shaded bands show the range of
the four scalar-rate prescriptions in Sec.~\ref{sec:scalar-inputs} for $\pi\pi$, $K\bar K$, and
$4\pi$, with each prescription's own total hadronic width in the
denominator. These are model envelopes without a statistical
confidence-level interpretation.
Above 2~GeV the two-meson rates follow the explicit amplitude
extrapolation in Eq.~\eqref{eq:scalar-two-meson-tail}; their widths are
subtracted before allocating the multihadron remainder.
The low-mass rates and their alternatives originate from
Refs.~\cite{Blackstone:2024ouf,Boiarska:2019jym,Winkler:2018qyg}.
\label{fig:sm-scalar-portal}}
\end{figure}

\begin{figure}[!htb]
\centering
\resizebox{\textwidth}{!}{%
\includegraphics{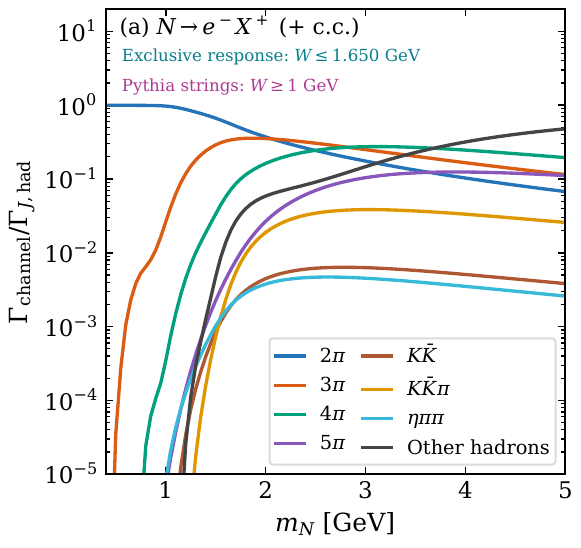}%
\includegraphics{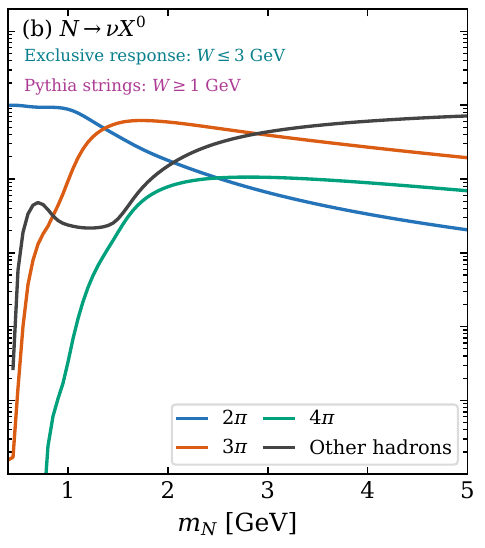}%
}
\caption{Channel fractions in nonstrange HNL multihadron decays through
the charged current (left) and neutral current (right), as functions of
the parent HNL mass, following Sec.~\ref{sec:hnl-inputs}.
Each panel is normalized to its own multihadron
current width. The probabilities at fixed hadronic mass $W$ use the
spectral input and tuned model; the weak decay spectrum is
then integrated over $W$ with its parent-mass-dependent weight.
One-meson modes and the remaining
flavor currents are separate contributions to the full HNL decay rate.
The spectral and weak-rate sources are
Refs.~\cite{ALEPH:2005qgp,Davier:2013sfa,Bondarenko:2018ptm}.
\label{fig:sm-hnl-portals}}
\end{figure}

Figs.~\ref{fig:sm-vector-portals}--\ref{fig:sm-hnl-portals} show how
the predicted hadronic decay rates are divided among channels.
For vectors, the ALP, and the scalar, each curve is
$\Gamma_F/\Gamma_{\rm had}$, including hadrons with photons but excluding
purely leptonic, invisible, and diphoton decays. The HNL panels
resolve the indicated nonstrange multihadron weak-current width,
averaged over the hadronic mass $W$ with the weak decay spectrum.
The accompanying lepton is excluded from the channel labels; one-meson
modes and other flavor currents contribute separately to the full HNL rate.
Tab.~\ref{tab:raw-comparisons} defines the generator references used
in these comparisons and in the SHiP analysis below.

\paragraph{Selection rules and neutral daughters.}
The $B-L$ current has no isovector component in the isospin limit, so
the two-pion rate is neglected. For the ALP, parity forbids the
two-pseudoscalar states populated by raw \textsc{Pythia}.
Its large $\eta^{(\prime)}\pi\pi$ and $K\bar K\pi$ contributions
feed higher-multiplicity states, often with photons from neutral-meson
decays. The $K\bar K\pi$ group includes $K^*\bar K$ followed by
$K^*\to K\pi$, but excludes two-kaon states.
For $B-L$, $\omega\pi\pi$ followed by $\omega\to3\pi$ feeds the five-pion
group, while $\eta+X$ combines $\eta3\pi$ and $\eta K\bar K$.
These daughter decays connect the channel fractions to the charged-only
and mixed signatures considered below.

\paragraph{Scalar and HNL composition.}
In Fig.~\ref{fig:sm-scalar-portal}, the large four-pion fraction near
2~GeV follows from assigning the light-hadronic width left after the
separately calculated two-meson and nucleon-pair modes to four pions.
At higher masses, the decreasing four-pion probability leaves a growing
multihadron remainder. The two-meson rates follow the amplitude continuation
and phase space of Eq.~\eqref{eq:scalar-two-meson-tail}.
The HNL fractions in Fig.~\ref{fig:sm-hnl-portals} average the channel
probabilities over the allowed hadronic masses $W$ at each parent mass $m_N$.
Features tied to a fixed hadronic
mass are therefore spread across the parent-mass dependence.

\paragraph{Remainder and plot conventions.}
As the known exclusive rates fall below the inclusive hadronic width,
an increasing fraction is assigned to unmeasured channels. Its total
rate follows from subtraction; its composition is modeled as specified
in Sec.~\ref{sec:portal-inputs}. Other collects all states without their
own curves and can include both calculated modes and this remainder.
The underlying categories are disjoint and sum to one; the captions
identify contributions omitted from the displayed curves.
Vector and ALP comparisons keep the total hadronic width fixed;
scalar bands also vary the rate calculation and its denominator.
Teal markers indicate total-width interpolation. Purple markers
indicate the start of channel matching for $B-L$ and the ALP, and
the start of remainder generation for the dark photon and scalar.
Tab.~\ref{tab:matching-map} specifies the corresponding intervals.

\begin{figure}[!htb]
\centering
\resizebox{\textwidth}{!}{%
\includegraphics{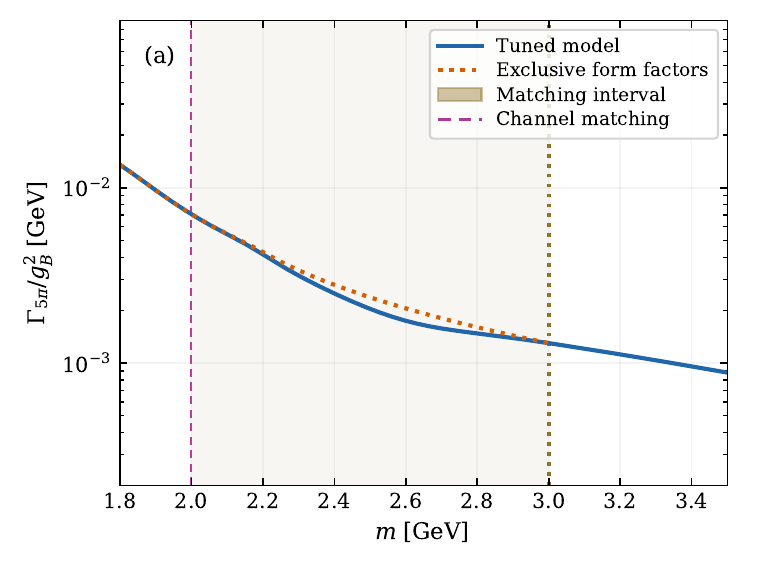}%
\includegraphics{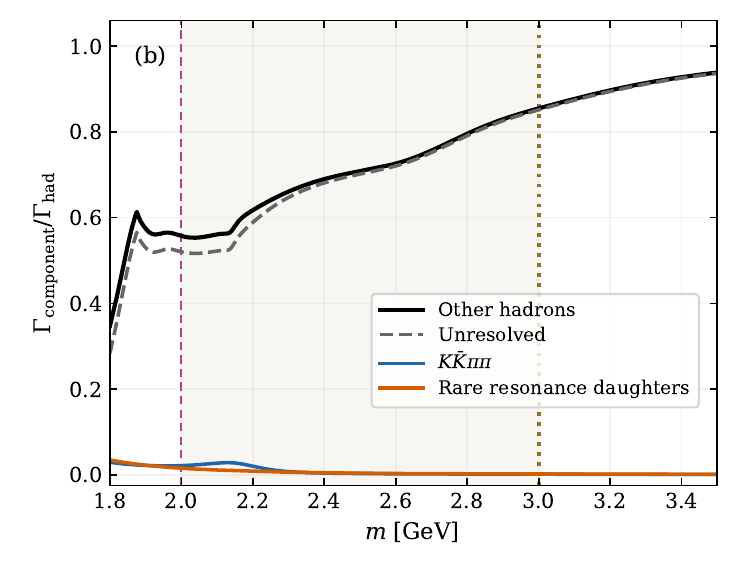}%
}
\caption{$B-L$ matching check. Left: the five-pion partial width from the
interpolation in Eq.~\eqref{eq:baryon-common-width} and the exclusive
form-factor continuation.
Above 2~GeV the dotted curve is a form-factor extrapolation used as input.
Right: the contributions to Other in our model. The
unresolved contribution is the inclusive width left after subtracting the
calculated channels. The small $\eta\gamma$ contribution is included in
Other but not drawn separately. Ochre marks the 2-3~GeV interpolation of
channel widths; purple marks the start of this interpolation.}
\label{fig:baryon-matching-audit}
\end{figure}

\clearpage
\paragraph{Resonance structure and matching features of the $B-L$ vector.}
\label{sec:bl-spectral-features}
Fig.~\ref{fig:baryon-matching-audit} contrasts the decreasing five-pion
width with the growing remainder; Fig.~\ref{fig:sm-vector-portals}
shows the full channel fractions.
The five-pion partial width is dominated by $\omega\pi\pi$ followed by
$\omega\to3\pi$. The form-factor fit of Ref.~\cite{Foguel:2022ppx} uses a single broad
$\omega(1650)$ resonance; BaBar measurements of $\omega\pi^+\pi^-$ and
$\omega\pi^0\pi^0$ support its rise and falling tail
\cite{BaBar:2007qju,BaBar:2018rkc,Foguel:2022ppx}.
This contribution accounts for more than nine tenths of $\Gamma_{5\pi}$
between 1.5 and 2.2~GeV. The smaller $\phi\pi\pi$ contribution includes
the $\phi(2170)$ resonance. Their sum has a smooth maximum near 1.7~GeV
and then decreases. Above the low-mass matching interval, the channel
fractions use the inclusive denominator in Eq.~\eqref{eq:baryon-inclusive}.

The kaon-pair form factors give a broad enhancement near 2.6~GeV from
constructive interference of the $\omega$- and $\phi$-like amplitudes
for the $B-L$ current. This predicted enhancement follows from the
flavor decomposition and its overlapping effective resonance terms.
Its uncertainty is separate from fragmentation.

Vector decays into nucleon pairs allow an $S$ wave. With the smooth
form factors used here, the width therefore opens as
$\beta_N=\sqrt{1-4m_N^2/m^2}\propto\sqrt{m-2m_N}$, where $m_N$ is the
nucleon mass. Its derivative is singular at threshold, producing a
sharp onset of the nucleon-pair fraction.
The calculation has no explicit Coulomb or nucleon-rescattering factor.

The narrow feature in $\eta+X$ near 2.14~GeV comes from the
$\eta\phi$ amplitude, followed by $\phi\to K\bar K$ or $3\pi$;
$\eta\omega$ is much smaller there. It is already present in the
electromagnetic fit of Ref.~\cite{Plehn:2019jeo}, whose Tab.~7 gives
a resonance width of $44\pm33$~MeV. The matching in
Eq.~\eqref{eq:baryon-common-width} slightly reduces this feature.
Its sharpness reflects the uncertainty of that resonance fit.

The turn of the five-pion fraction near 1.75~GeV in
Fig.~\ref{fig:branchings} comes from matching the exclusive total hadronic width
to the inclusive width over 1.70--1.78~GeV
(Sec.~\ref{sec:hadronic-composition}). Its partial width follows the smooth resonance form factors
throughout this interval. The precise position and curvature of the
turn depend on the total-width interpolation.

\clearpage
\subsection{Model variations and uncertainty coverage}
\label{sec:uncertainties}

We quantify individual sources of uncertainty separately.
The ALP bands in Fig.~\ref{fig:ship} propagate the same fragmentation
variations through charged-only and mixed charged-photon selections.
Tab.~\ref{tab:ship-numerical} separates their model intervals from Monte Carlo errors.
Scalar bands compare
exclusive-rate calculations. Matching choices and uncertainties in
transferring inputs between currents are not generally covered.

\paragraph{Fragmentation and the mass dependence.}
The ALP SHiP comparison includes 17 choices: the central probability
prescription in Eq.~\eqref{eq:portable-law}, the alternative
$T_F=1$ in that equation, two alternative exponents $p=1.75,2.25$,
and 13 variations of the \textsc{Pythia} fragmentation parameters.
The fragmentation variations change the stopping mass, the longitudinal
fragmentation parameters, the transverse momentum width, or the
$e^+e^-$ tune. The stopping-mass alternatives are 0.1, 0.2, 0.4, and
0.8~GeV, compared with zero in the central calculation. Each choice is
propagated through the decay events and both detector selections.
The plotted envelope is the pointwise minimum and maximum of these
calculations, with the matching mass and the channel probabilities at
that mass held fixed.

\paragraph{Exclusive rates and transfer between currents.}
Experimental and form-factor errors affect the exclusive input rates.
For $B-L$, these include the local
electromagnetic $3\pi$ fit and the relative flavor amplitudes.
Fig.~10 of Ref.~\cite{Ilten:2018crw} shows mass-dependent uncertainties
in a vector model with the same hadronic current; different leptonic
channels change its total-width denominator. That study also estimates
a roughly twenty-percent uncertainty from vector-meson dominance and
flavor symmetry. Fig.~7 of Ref.~\cite{Foguel:2022ppx} illustrates the
spread between exclusive $B-L$ calculations. Our local fit varies the
electromagnetic normalization and slope while keeping the flavor
amplitudes fixed. These input uncertainties are not propagated into a
combined $B-L$ band: Figs.~\ref{fig:branchings} and
\ref{fig:sm-vector-portals} show central predictions.
The ALP and scalar rates depend
on their meson mixing and hadronic form factors. For the scalar,
Fig.~\ref{fig:sm-scalar-portal} shows the spread of the four scalar-rate
prescriptions defined above for $\pi\pi$, $K\bar K$, and $4\pi$, propagated
through Eq.~\eqref{eq:scalar-two-meson-tail}. This includes their
correlated change in the total hadronic width. Uncertainty in the assumed
high-mass dependence of the two-meson amplitudes remains outside this band.
For dark photons, the experimental comparison points show pointwise
measurement errors. Uncertainties and correlations in the input cross
sections have not been propagated into the dark-photon channel fractions
or SHiP accepted fractions shown here. The dark-photon errors in
Tab.~\ref{tab:ship-numerical} are Monte Carlo errors.
Hadronization variations were not generated for the dark-photon samples.
The unmeasured remainder is 40\% of hadronic decays at 2~GeV and 75\%
at 3~GeV. Within it, the charge composition of all-pion states is fixed
by the statistical-isospin rule of Sec.~\ref{sec:dark-photon-inputs}, so
the variations would mainly change the multiplicity distribution and the
kaon content. We expect an effect of similar size to the ALP envelope;
it has not been computed.

\paragraph{Matching and incomplete channel information.}
The interpolation intervals and the channel probabilities at the
matching mass are further model assumptions. Additional resonances
absent from the exclusive inputs can change both the widths and the
channel probabilities. Variations of the fragmentation parameters and
the power-law exponent do not estimate these effects.
Uncalculated exclusive
widths affect the division of the inclusive $B-L$ width among channels.
Perturbative and low-mass matching uncertainties belong to the separate
total-width calculation. The ALP envelope does not include uncertainties
in the exclusive input's division of the $K\bar K\pi$ rate among charge
channels.

\subsection{Accepted signatures at SHiP}
\label{sec:events-and-ship}

We compare the fractions of hadronic decays whose final particles all
pass the SHiP geometric and momentum selections, either as charged-only states or
as mixed charged-photon states.

\paragraph{Observable.}
For each category $\mathcal C$, we repeat Eq.~\eqref{eq:ship-acceptance}
from the main text:
\begin{equation}
f_{\mathcal C}^{\mathrm{rec}} =
\frac{N_{\mathcal C}^{\text{all }n\text{ particles accepted}}}
{N_{\mathrm{had}}^{\mathrm{in\ volume}}}.
\label{eq:sm-ship-acceptance}
\end{equation}
Here $n$ counts final particles after daughter decays. The numerator
counts accepted decays in category $\mathcal C$; the denominator counts
all hadronic decays in the volume. Both include production, branching
fractions, and decay-probability weights. The fraction therefore includes
both the probability of producing the selected state and its selection
acceptance.
For dark photons, both sums include production in the initial
proton-target collision and in subsequent interactions of secondary
particles in the target.

\paragraph{Selections.}
Two charged-only selections are applied after the daughter decays included in the
event generator: exactly two final particles, both charged; or at least
four final particles, all charged. Every particle must also cross the
charged-particle analysis plane. A state containing two charged particles and any
additional neutral particles passes neither charged-only selection.
The green curves select mixed states with all final particles accepted:
at least two charged particles, at least one photon, and no other
final particles after daughter decays. Every charged particle and photon
must pass the corresponding geometric selection and the momentum cut specified below.
Events containing neutrinos, neutrons, or $K_L^0$
fail this selection. The three categories are disjoint, so their accepted
fractions can be added to obtain the contribution from their union.

\paragraph{Geometry and reconstruction assumptions.}
The proper decay length is $c\tau=10$~m. In the geometry of
Refs.~\cite{SHiP:2018xqw,SHiP:2020vbd}, $z$ is the distance along the beam
from the target center. The decay volume extends from
$z=32$ to 82~m, widening linearly from $1\times2.7$~m$^2$ to
$4\times6.2$~m$^2$. The calculation in Fig.~\ref{fig:ship} and
Tab.~\ref{tab:ship-numerical} approximates the downstream detector
coverage by centered $4\times6$~m$^2$ planes at the spectrometer end,
$z=95$~m, for charged particles and at the electromagnetic-calorimeter
end, $z=97.32$~m, for photons. Every final particle must have laboratory
three-momentum $|\mathbf p|>1$~GeV; for photons this is equivalent to
$E_\gamma>1$~GeV. Particle trajectories are straight and reconstruction
efficiency is set to unity. Calorimeter shower resolution and backgrounds
are outside this comparison. These cuts affect only the numerator of
Eq.~\eqref{eq:sm-ship-acceptance}; its denominator includes all hadronic
decays inside the volume.

\paragraph{Comparison setup.}
\textsc{EventCalc-SHiP}~\cite{EventCalc} uses the same parent production and
decay positions for both descriptions at each mass, with the same
proper lifetime and total hadronic branching fraction. For both dark
photons and ALPs, our model is compared with raw \textsc{Pythia} for the entire hadronic decay
component. The ALP comparison
therefore includes both the selection rules and the changes in the
probabilities of allowed channels.

Both particle comparisons cover 1.5--3~GeV, the range of the SHiP event
samples used for this low-mass benchmark. The decay model extends to
higher masses, as shown in the channel-fraction plots above.
Below 1.7~GeV, the dark-photon
model prediction uses only the exclusive electromagnetic calculation.
The dark-photon raw comparator is displayed only at and above the
inclusive-switching mass of 1.7~GeV. The scope of the errors and bands
in Fig.~\ref{fig:ship} is specified in Sec.~\ref{sec:uncertainties}.
Tab.~\ref{tab:ship-numerical} reports the fractions at 2~GeV.

Below the ALP matching mass, the channel probabilities and momentum
distributions follow the exclusive calculation of
Ref.~\cite{Ovchynnikov:2025gpx}. The raw comparison also extends to
1.5~GeV: separately
generated $gg$ and $s\bar s$ samples use the source probabilities
of approximately $0.89$ and $0.11$ at $\approx1.9$~GeV, held fixed below that mass.
These weights define the low-mass raw comparison. Our model uses the exclusive
calculation below the matching mass. The fragmentation-variation band starts
at that mass.

\paragraph{Acceptance near the matching masses.}
\label{sec:ship-momentum-boundary}
We test how the choice of daughter-momentum generator near the matching
mass affects the mixed charged-photon selection defined above.
For the dominant ALP channel $a\to\eta\pi^+\pi^-$ at 1.911~GeV,
the acceptance is the fraction of all decays in this channel inside the
volume that pass this selection, including the parent production and
decay-probability weights. We compare momenta generated with the exclusive
matrix element of Ref.~\cite{Ovchynnikov:2025gpx} with \textsc{Pythia}
fragmentation events selected to have the same channel.
Each pair has identical daughter masses, parent momenta and decay vertices,
and uses the same sampled decay modes and rest-frame kinematics of the
daughter particles.

We repeat the integration with new exclusive phase-space points and
daughter decays, and a different pairing with the same parent events.
The differences, defined as \textsc{Pythia} minus exclusive acceptance,
are $-0.94\pm0.14$ and $-0.72\pm0.14$ percentage points in the two
integrations, respectively. These correspond to decreases of 6\% and 5\%
relative to the exclusive acceptance. The errors are Monte Carlo standard
errors. Both integrations use the same 55,359 stored \textsc{Pythia}
primary-momentum configurations, so their estimates are correlated.
\textsc{Pythia} fragmentation supplies its own momentum correlations,
which can differ from constant-matrix-element phase space.

For dark photons, the accepted fraction for mixed charged-photon events,
normalized to all hadronic decays inside the volume as in
Eq.~\eqref{eq:sm-ship-acceptance}, is 18.43\% at 1.699~GeV and
18.88\% at 1.700~GeV, a relative increase of about 2.4\%.
The individual Monte Carlo errors are about 0.13 percentage points.
The exclusive channel fractions follow the same form-factor calculation
across this boundary, where generation of the unmeasured remainder with
\textsc{Pythia} starts (Sec.~\ref{sec:dark-photon-inputs}).
The two points use different routines for generating primary momenta and
decaying daughter particles, and separate Monte Carlo event samples.
The size and origin of acceptance differences can therefore depend on
the channel, particle, and experimental selection.

The green curves and the dashed-blue ALP curve in Fig.~\ref{fig:ship} are
smoothed visual guides; numerical comparisons use the original Monte Carlo
estimates. Sec.~\ref{sec:ship-curve-display} gives the plotting prescription.

\paragraph{Results and limitations.}
For the ALP benchmark with universal fermion couplings at
$\Lambda=1$~TeV, the accepted fraction with exactly two charged particles is
suppressed at every mass where the raw comparison is available and for
every model variation.
The fraction with at least four charged particles changes less
uniformly: the central result is generally lower than raw \textsc{Pythia},
while the stopping-mass variation that uses the \textsc{Pythia} default
approaches the raw result in the middle of the simulated range.
Near its upper end, the central result is also close to raw \textsc{Pythia}.
For dark photons in the 2--3~GeV region, the exactly-two fraction decreases
and the at-least-four fraction increases relative to raw \textsc{Pythia}.

For the ALP, the exclusive calculation gives no fully charged two-particle
decays below the proton-antiproton threshold. Its small fully charged
higher-multiplicity contribution includes
four-pion states, charged-kaon modes with $K_S^0\to\pi^+\pi^-$, and
small fully charged decay branches of the $\eta^{(\prime)}$ mesons.

Mixed charged-photon states give a larger accepted fraction than either
charged-only category for both particles at 2~GeV
(Tab.~\ref{tab:ship-numerical}). Including them therefore changes which
signatures dominate the accepted population and motivates combined
tracking and calorimetry.

% Numerical values remain tied to the authenticated EventCalc analysis.
% BEGIN AUTHENTICATED SHIP VALUES
% Reanalysed from authenticated events with z_charged=95 m,
% z_photon=97.32 m and laboratory |p|>1 GeV for every final particle.
% Source: exhad/plots/model1_ship_topologies_isospin_v2.json.
% *Fraction macros are unitless numbers in [0,1].
% *Percent macros are numeric percentages without a percent sign.
\newcommand{\ShipTopologyReferenceMassGeV}{2.0}
\newcommand{\ShipMatchedEMDPExactlyTwoCentralVariant}{matched-em}
\newcommand{\ShipMatchedEMDPExactlyTwoRawFraction}{0.1076}
\newcommand{\ShipMatchedEMDPExactlyTwoRawPercent}{10.76}
\newcommand{\ShipMatchedEMDPExactlyTwoRawMCSEFraction}{0.004267}
\newcommand{\ShipMatchedEMDPExactlyTwoRawMCSEPercent}{0.43}
\newcommand{\ShipMatchedEMDPExactlyTwoCentralFraction}{0.0176}
\newcommand{\ShipMatchedEMDPExactlyTwoCentralPercent}{1.76}
\newcommand{\ShipMatchedEMDPExactlyTwoCentralMCSEFraction}{0.001864}
\newcommand{\ShipMatchedEMDPExactlyTwoCentralMCSEPercent}{0.19}
\newcommand{\ShipMatchedEMDPExactlyTwoEnvelopeLowFraction}{0.0176}
\newcommand{\ShipMatchedEMDPExactlyTwoEnvelopeLowPercent}{1.76}
\newcommand{\ShipMatchedEMDPExactlyTwoEnvelopeHighFraction}{0.0176}
\newcommand{\ShipMatchedEMDPExactlyTwoEnvelopeHighPercent}{1.76}
\newcommand{\ShipMatchedEMDPExactlyTwoEnvelopeMinusFraction}{0.0000}
\newcommand{\ShipMatchedEMDPExactlyTwoEnvelopeMinusPercent}{0.00}
\newcommand{\ShipMatchedEMDPExactlyTwoEnvelopePlusFraction}{0.0000}
\newcommand{\ShipMatchedEMDPExactlyTwoEnvelopePlusPercent}{0.00}
\newcommand{\ShipMatchedEMDPExactlyTwoEnvelopeApplicable}{0}
\newcommand{\ShipMatchedEMDPExactlyTwoRaw}{\ShipMatchedEMDPExactlyTwoRawFraction}
\newcommand{\ShipMatchedEMDPExactlyTwoPrediction}{\ShipMatchedEMDPExactlyTwoCentralFraction}
\newcommand{\ShipMatchedEMDPExactlyTwoEnvelopeLow}{\ShipMatchedEMDPExactlyTwoEnvelopeLowFraction}
\newcommand{\ShipMatchedEMDPExactlyTwoEnvelopeHigh}{\ShipMatchedEMDPExactlyTwoEnvelopeHighFraction}
\newcommand{\ShipMatchedEMDPFourOrMoreCentralVariant}{matched-em}
\newcommand{\ShipMatchedEMDPFourOrMoreRawFraction}{0.0256}
\newcommand{\ShipMatchedEMDPFourOrMoreRawPercent}{2.56}
\newcommand{\ShipMatchedEMDPFourOrMoreRawMCSEFraction}{0.001978}
\newcommand{\ShipMatchedEMDPFourOrMoreRawMCSEPercent}{0.20}
\newcommand{\ShipMatchedEMDPFourOrMoreCentralFraction}{0.0828}
\newcommand{\ShipMatchedEMDPFourOrMoreCentralPercent}{8.28}
\newcommand{\ShipMatchedEMDPFourOrMoreCentralMCSEFraction}{0.003554}
\newcommand{\ShipMatchedEMDPFourOrMoreCentralMCSEPercent}{0.36}
\newcommand{\ShipMatchedEMDPFourOrMoreEnvelopeLowFraction}{0.0828}
\newcommand{\ShipMatchedEMDPFourOrMoreEnvelopeLowPercent}{8.28}
\newcommand{\ShipMatchedEMDPFourOrMoreEnvelopeHighFraction}{0.0828}
\newcommand{\ShipMatchedEMDPFourOrMoreEnvelopeHighPercent}{8.28}
\newcommand{\ShipMatchedEMDPFourOrMoreEnvelopeMinusFraction}{0.0000}
\newcommand{\ShipMatchedEMDPFourOrMoreEnvelopeMinusPercent}{0.00}
\newcommand{\ShipMatchedEMDPFourOrMoreEnvelopePlusFraction}{0.0000}
\newcommand{\ShipMatchedEMDPFourOrMoreEnvelopePlusPercent}{0.00}
\newcommand{\ShipMatchedEMDPFourOrMoreEnvelopeApplicable}{0}
\newcommand{\ShipMatchedEMDPFourOrMoreRaw}{\ShipMatchedEMDPFourOrMoreRawFraction}
\newcommand{\ShipMatchedEMDPFourOrMorePrediction}{\ShipMatchedEMDPFourOrMoreCentralFraction}
\newcommand{\ShipMatchedEMDPFourOrMoreEnvelopeLow}{\ShipMatchedEMDPFourOrMoreEnvelopeLowFraction}
\newcommand{\ShipMatchedEMDPFourOrMoreEnvelopeHigh}{\ShipMatchedEMDPFourOrMoreEnvelopeHighFraction}
\newcommand{\PortableModelOneShipALPExactlyTwoCentralVariant}{central}
\newcommand{\PortableModelOneShipALPExactlyTwoRawFraction}{0.0834}
\newcommand{\PortableModelOneShipALPExactlyTwoRawPercent}{8.34}
\newcommand{\PortableModelOneShipALPExactlyTwoRawMCSEFraction}{0.003419}
\newcommand{\PortableModelOneShipALPExactlyTwoRawMCSEPercent}{0.34}
\newcommand{\PortableModelOneShipALPExactlyTwoCentralFraction}{0.0058}
\newcommand{\PortableModelOneShipALPExactlyTwoCentralPercent}{0.58}
\newcommand{\PortableModelOneShipALPExactlyTwoCentralMCSEFraction}{0.001047}
\newcommand{\PortableModelOneShipALPExactlyTwoCentralMCSEPercent}{0.10}
\newcommand{\PortableModelOneShipALPExactlyTwoEnvelopeLowFraction}{0.0039}
\newcommand{\PortableModelOneShipALPExactlyTwoEnvelopeLowPercent}{0.39}
\newcommand{\PortableModelOneShipALPExactlyTwoEnvelopeHighFraction}{0.0075}
\newcommand{\PortableModelOneShipALPExactlyTwoEnvelopeHighPercent}{0.75}
\newcommand{\PortableModelOneShipALPExactlyTwoEnvelopeMinusFraction}{0.0018}
\newcommand{\PortableModelOneShipALPExactlyTwoEnvelopeMinusPercent}{0.18}
\newcommand{\PortableModelOneShipALPExactlyTwoEnvelopePlusFraction}{0.0017}
\newcommand{\PortableModelOneShipALPExactlyTwoEnvelopePlusPercent}{0.17}
\newcommand{\PortableModelOneShipALPExactlyTwoEnvelopeApplicable}{1}
\newcommand{\PortableModelOneShipALPExactlyTwoRaw}{\PortableModelOneShipALPExactlyTwoRawFraction}
\newcommand{\PortableModelOneShipALPExactlyTwoPrediction}{\PortableModelOneShipALPExactlyTwoCentralFraction}
\newcommand{\PortableModelOneShipALPExactlyTwoEnvelopeLow}{\PortableModelOneShipALPExactlyTwoEnvelopeLowFraction}
\newcommand{\PortableModelOneShipALPExactlyTwoEnvelopeHigh}{\PortableModelOneShipALPExactlyTwoEnvelopeHighFraction}
\newcommand{\PortableModelOneShipALPFourOrMoreCentralVariant}{central}
\newcommand{\PortableModelOneShipALPFourOrMoreRawFraction}{0.0190}
\newcommand{\PortableModelOneShipALPFourOrMoreRawPercent}{1.90}
\newcommand{\PortableModelOneShipALPFourOrMoreRawMCSEFraction}{0.001594}
\newcommand{\PortableModelOneShipALPFourOrMoreRawMCSEPercent}{0.16}
\newcommand{\PortableModelOneShipALPFourOrMoreCentralFraction}{0.0059}
\newcommand{\PortableModelOneShipALPFourOrMoreCentralPercent}{0.59}
\newcommand{\PortableModelOneShipALPFourOrMoreCentralMCSEFraction}{0.000917}
\newcommand{\PortableModelOneShipALPFourOrMoreCentralMCSEPercent}{0.09}
\newcommand{\PortableModelOneShipALPFourOrMoreEnvelopeLowFraction}{0.0042}
\newcommand{\PortableModelOneShipALPFourOrMoreEnvelopeLowPercent}{0.42}
\newcommand{\PortableModelOneShipALPFourOrMoreEnvelopeHighFraction}{0.0065}
\newcommand{\PortableModelOneShipALPFourOrMoreEnvelopeHighPercent}{0.65}
\newcommand{\PortableModelOneShipALPFourOrMoreEnvelopeMinusFraction}{0.0017}
\newcommand{\PortableModelOneShipALPFourOrMoreEnvelopeMinusPercent}{0.17}
\newcommand{\PortableModelOneShipALPFourOrMoreEnvelopePlusFraction}{0.0006}
\newcommand{\PortableModelOneShipALPFourOrMoreEnvelopePlusPercent}{0.06}
\newcommand{\PortableModelOneShipALPFourOrMoreEnvelopeApplicable}{1}
\newcommand{\PortableModelOneShipALPFourOrMoreRaw}{\PortableModelOneShipALPFourOrMoreRawFraction}
\newcommand{\PortableModelOneShipALPFourOrMorePrediction}{\PortableModelOneShipALPFourOrMoreCentralFraction}
\newcommand{\PortableModelOneShipALPFourOrMoreEnvelopeLow}{\PortableModelOneShipALPFourOrMoreEnvelopeLowFraction}
\newcommand{\PortableModelOneShipALPFourOrMoreEnvelopeHigh}{\PortableModelOneShipALPFourOrMoreEnvelopeHighFraction}
% END AUTHENTICATED SHIP VALUES
% Model mixed-state values from the 2-GeV mixed_prediction entries in
% exhad/plots/model1_ship_visible_reconstruction_isospin_v2.json.
\newcommand{\ShipMixedDPCentralPercent}{14.34}
\newcommand{\ShipMixedDPMCSEPercent}{0.44}
\newcommand{\ShipMixedALPCentralPercent}{8.96}
\newcommand{\ShipMixedALPMCSEPercent}{0.33}
\newcommand{\ShipMixedALPEnvelopeLowPercent}{8.63}
\newcommand{\ShipMixedALPEnvelopeHighPercent}{9.30}
\begin{table*}[t!]
\centering
\begin{tabular}{llr@{\hspace{1em}}r@{\hspace{1em}}c}
\toprule
Particle & Final state & Raw \textsc{Pythia} & Our model & Model interval\\
\midrule
Dark photon & Exactly two charged &
$\ShipMatchedEMDPExactlyTwoRawPercent\pm\ShipMatchedEMDPExactlyTwoRawMCSEPercent$ &
$\ShipMatchedEMDPExactlyTwoCentralPercent\pm\ShipMatchedEMDPExactlyTwoCentralMCSEPercent$ & ---\\
 & At least four, all charged &
$\ShipMatchedEMDPFourOrMoreRawPercent\pm\ShipMatchedEMDPFourOrMoreRawMCSEPercent$ &
$\ShipMatchedEMDPFourOrMoreCentralPercent\pm\ShipMatchedEMDPFourOrMoreCentralMCSEPercent$ & ---\\
 & Mixed tracks-photons & --- &
$\ShipMixedDPCentralPercent\pm\ShipMixedDPMCSEPercent$ & ---\\
ALP & Exactly two charged &
$\PortableModelOneShipALPExactlyTwoRawPercent\pm\PortableModelOneShipALPExactlyTwoRawMCSEPercent$ &
$\PortableModelOneShipALPExactlyTwoCentralPercent\pm\PortableModelOneShipALPExactlyTwoCentralMCSEPercent$ &
$[\PortableModelOneShipALPExactlyTwoEnvelopeLowPercent,
\PortableModelOneShipALPExactlyTwoEnvelopeHighPercent]$\\
 & At least four, all charged &
$\PortableModelOneShipALPFourOrMoreRawPercent\pm\PortableModelOneShipALPFourOrMoreRawMCSEPercent$ &
$\PortableModelOneShipALPFourOrMoreCentralPercent\pm\PortableModelOneShipALPFourOrMoreCentralMCSEPercent$ &
$[\PortableModelOneShipALPFourOrMoreEnvelopeLowPercent,
\PortableModelOneShipALPFourOrMoreEnvelopeHighPercent]$\\
 & Mixed tracks-photons & --- &
$\ShipMixedALPCentralPercent\pm\ShipMixedALPMCSEPercent$ &
$[\ShipMixedALPEnvelopeLowPercent,\ShipMixedALPEnvelopeHighPercent]$\\
\bottomrule
\end{tabular}
\caption{SHiP accepted fractions of Eq.~\eqref{eq:sm-ship-acceptance}, in percent,
at $m=\ShipTopologyReferenceMassGeV$~GeV and $c\tau=10$~m.
Charged-only rows accept no neutral particles; mixed states require at
least two charged particles and at least one photon, with every final
particle either charged or a photon.
Every particle must have laboratory momentum $|\mathbf p|>1$~GeV and
cross its detector plane: $z=95$~m for charged particles and 97.32~m
for photons (Sec.~\ref{sec:events-and-ship}).
Errors are Monte Carlo standard errors. The last column spans the ALP
hadronization-model variations defined in Sec.~\ref{sec:uncertainties};
a dash denotes an unevaluated interval. Raw mixed-state fractions are
not tabulated. Dark-photon data and hadronization-model uncertainties are not
included in the quoted errors.
\label{tab:ship-numerical}}
\end{table*}

\section{Implementation and numerical details}
\label{sec:implementation}

\subsection{Standalone decay generation and simulation interfaces}
\exhad~\cite{exhad} is a standalone decay package for experimental
frameworks using \textsc{Pythia}; it does not depend on
\textsc{EventCalc-SHiP}. Its numerical inputs contain the decay rates,
channel definitions, and coefficients needed to evaluate the model at
the requested mass. Input preparation and fitting are performed separately
from event generation. Given a particle model, mass, and random seed,
\exhad selects channels with the prescribed probabilities and returns
daughter identities and four-momenta in the parent rest frame. Separately
generated modes are excluded from fragmentation to avoid double counting.

The package provides Python and C++ interfaces, including a
\textsc{Pythia} external-decay handler. The C++ interface uses a persistent
Python controller for the compiled generators. Events can also be written as JSON
or HepMC records containing the parent and final daughters. Short-lived
secondary particles are already decayed; long-lived daughters remain
available for detector transport.

Boson samples can be restricted to hadronic decays or include the full set
of hadronic and nonhadronic modes supplied by the input branching
fractions. Full HNL decays additionally use the three squared mixing
amplitudes. For a hadronic-only boson sample, the simulation includes the
total hadronic branching fraction once; a full-decay sample already
includes this channel selection. Parent production, lifetime, decay position,
and detector response belong to the simulation using \exhad.

\textsc{EventCalc-SHiP}~\cite{EventCalc} is one such application, used here
for the SHiP comparison. It samples parent production and decay positions, boosts the
daughters to the laboratory, and applies the geometric and momentum selections in
Sec.~\ref{sec:events-and-ship}. It combines production and decay weights
with the total hadronic branching fraction, counted once. Event
reconstruction efficiency is set to unity. Detector response and
backgrounds are not simulated.

For HNLs, the weak decay fixes the
hadronic invariant mass $W$ and the accompanying lepton kinematics;
the hadronic generator is called at that $W$, with the specified weak
current. The selected-current distributions in
Fig.~\ref{fig:sm-hnl-portals} are combined with the separately calculated
one-meson and other flavor contributions when constructing full HNL
decays.

\paragraph{HNL and ALP mass ranges.}
\label{sec:hnl-numerical-range}
The standalone HNL decay tables span $0.02\leq m_N\leq40$~GeV.
This range is independent of the parent
production spectra available in a particular experiment simulation.
For the fermion-universal ALP, the input table ends at 5.2~GeV.
For both particles, the low-mass hadronic description is joined to
showered \textsc{Pythia} over 4--5~GeV: the relevant mass is $W$ for an
HNL and the parent mass for an ALP. In this interval, an event uses the
showered generator with probability $h(x)=10x^3-15x^4+6x^5$, where
$x$ is the relevant mass minus 4~GeV, divided by 1~GeV.
Below the interval $h=0$, and above it $h=1$.
This is a modelling prescription for the event distribution: it selects
one complete generator, without averaging daughter momenta or changing
the supplied partial widths. HNL heavy-flavor currents without a
low-mass description use showered \textsc{Pythia} above their hadronic
thresholds. The ALP quantum-number restrictions remain active before
secondary decays; in particular, $D\bar D$ is forbidden, whereas
$D\bar D^*$ and its charge conjugate are allowed.

\subsection{Sampling and statistical estimators}
\label{sec:statistical-estimators}

Ordinary generation returns unweighted decays. For ALP
$\eta^{(\prime)}\pi\pi$ channels at $2.4<m\leq3.5$~GeV, a conditional
sampler removes incompatible fragmentation histories while preserving the
probabilities and momentum distributions of accepted events.
A separate weighted ALP option instead attaches importance weights
that must enter all observables. Replacing event-level vetoes by retuned
local fragmentation requires preserving both channel probabilities and
the momentum distributions within channels.

For the pion-remainder generation in Sec.~\ref{sec:dark-photon-inputs}, the
\textsc{EventCalc-SHiP} interface uses a limit of 65,536 \textsc{Pythia}
trials to obtain each selected charge configuration. The parameter
\texttt{maximum\_reference\_condition\_attempts} sets this limit;
exhausting it invokes
the flat-phase-space fallback described there.

The SHiP dark-photon samples contain 100,000 parent trials per production
source at each of the five masses from 1.65 to 1.71~GeV and 10,000 at
the other masses. The nine ALP masses below $\approx1.9$~GeV each use 200,000
parent trials. Raw \textsc{Pythia} and our model share the parent samples;
production and decay probabilities enter as weights.

For each independent parent trial $j$, let $\omega_j$ include the
production normalization, hadronic branching fraction, and decay-probability
weight, and let $a_{j\mathcal C}$ equal one if the event passes selection
$\mathcal C$ and zero otherwise. The accepted fraction and its Monte Carlo
standard error are
\begin{equation}
 f_{\mathcal C}^{\rm rec}=\frac{\sum_j\omega_j a_{j\mathcal C}}{\sum_j\omega_j},
 \qquad
 \delta f_{\mathcal C}^{\rm rec}=
 \frac{\sqrt{\sum_j\omega_j^2(a_{j\mathcal C}-f_{\mathcal C}^{\rm rec})^2}}
 {\sum_j\omega_j}.
 \label{eq:ship-mc-error}
\end{equation}
For raw flavor samples sharing the same parent trials, $a_{j\mathcal C}$
is the source-probability-weighted mean of their selection indicators.
Combining these indicators before estimating the error accounts for the
shared parent sample.

For the boson benchmarks other than the electromagnetic test, a half-count
per allowed category stabilizes the estimated probabilities:
$(k_{F,a}+1/2)/(N_a+K_a/2)$, where $k_{F,a}$ is the count among $N_a$
events and $K_a$ counts the allowed categories, including the remainder.
Forbidden categories stay at zero. HNLs use direct counts. The
electromagnetic test averages direct counts from five runs; its
uncertainty on the rise at the first mass point includes the fitted-power
error. For the other applications, the errors used to test the finite-mass rise are
Monte Carlo counting errors on the channel probabilities $Q^{\rm Py}_{F,a}$
estimated separately from the generated events for each source $a$.

The raw-generator bands in Figs.~\ref{fig:raw-failures} and
\ref{fig:em-repair} represent Monte Carlo counting errors.
For a zero count, the conservative 95\% upper bound on the channel
probability is $-\ln(0.05)\max_a(r_a/N_a)$, where $N_a$ is the number
of independent events from source $a$; the cross-section limit multiplies
this bound by the same inclusive light-quark cross section as the curve.

\paragraph{Curves in the SHiP comparison.}
\label{sec:ship-curve-display}
The green curves and the dashed-blue ALP curve in Fig.~\ref{fig:ship} use
cubic smoothing splines weighted by inverse Monte Carlo errors, with a sum
of squared standardized residuals bounded by the number of mass points.
For the dashed-blue ALP curve, shape-preserving interpolation between
the fitted values avoids overshoot between sampled masses.
The other blue and orange curves use shape-preserving cubic interpolation
through the calculated values. In our model, the ALP fraction with exactly
two charged particles remains zero below the proton-antiproton production
threshold.
The blue and orange ALP bands enclose all interpolated systematic variations
and retain their original bounds at the sampled masses. The green band
encloses the same variations after applying the green-curve smoothing to each.

\subsection{Interpolation coefficients and numerical boundaries}
\label{sec:interpolation-details}

For the ALP, we interpolate the vector $\mathbf P$ containing
all $P_F$ and the remainder $P_R$. Let $\mathbf b$ and $\mathbf b'$ be
its exclusive value and mass derivative at $\mstar$, and
$\mathbf p_1$ and $\mathbf p_1'$ the value and derivative of
Eq.~\eqref{eq:portable-law} at $m_1=3$~GeV. Define
$L=m_1-\mstar$ and $x=(m-\mstar)/L$. The numerical prescription is
\begin{equation}
 \mathbf P(m)=\sum_{j=0}^{5}\binom{5}{j}
 x^j(1-x)^{5-j}\mathbf c_j ,
 \label{eq:alp-probability-match}
\end{equation}
with $\mathbf c_0=\mathbf b$,
$\mathbf c_1=\mathbf b+L\mathbf b'/5$,
$\mathbf c_4=\mathbf p_1-L\mathbf p_1'/5$, and
$\mathbf c_5=\mathbf p_1$. We fix the remaining coefficients as
$\mathbf c_2=(2\mathbf c_1+\mathbf c_4)/3$ and
$\mathbf c_3=(\mathbf c_1+2\mathbf c_4)/3$.
Every coefficient vector is required to be nonnegative and to sum to one,
which ensures nonnegative, normalized probabilities throughout the interval.
The endpoint coefficients match the values and first derivatives at both ends.

Tab.~\ref{tab:numerical-boundaries} records the short numerical
connections at the edges of the exclusive input grids.

\begin{table*}[!htbp]
\centering
\begin{tabular}{p{0.23\textwidth}p{0.19\textwidth}p{0.47\textwidth}}
\toprule
Quantity & Mass or interval & Prescription\\
\midrule
ALP exclusive probabilities & $\mstar=1.911$ &
Backward-difference slopes over the last 5~MeV; linear continuation of
the last exclusive value by 1~MeV to the fragmentation boundary\\
ALP nucleon-pair widths & 1.910-1.915 &
Monotone cubic Hermite interpolation of the separate absolute widths\\
ALP source fractions & $\mstar$--1.915 &
Hold the nonhadronic branching fractions and $gg:s\bar s$ ratio at
their 1.915~GeV values. Normalize the $gg$ and $s\bar s$ sum to the
hadronic width minus the separate nucleon-pair widths, using the
mass-dependent total width\\
HNL charged-current probability curves & $W=1.650$-1.651 &
Linear connection from the exclusive boundary to the first generated point\\
Scalar $\pi\pi$, $K\bar K$, $4\pi$ widths & 1.999-2.000 &
Continue two-meson amplitudes with Eq.~\eqref{eq:scalar-two-meson-tail};
assign the remaining light-hadronic width to $4\pi$ below the generator switch\\
\bottomrule
\end{tabular}
\caption{Numerical treatment at the boundaries of the exclusive inputs
for the prescriptions in Sec.~\ref{sec:portal-inputs}; the ALP probability
interpolation is given in Eq.~\eqref{eq:alp-probability-match}.
The broader physical matching intervals are given in
Tab.~\ref{tab:matching-map}. Masses and intervals are in GeV.}
\label{tab:numerical-boundaries}
\end{table*}

\subsection{Event-sample sizes and consistency checks}
The source samples used to determine the channel probabilities contain
$6.2\times10^8$ events for the central settings and $3.1\times10^8$ events
combined across all 15 fragmentation-parameter variations. Each total sums
over the quark and gluon sources, current components, masses, and independent
runs; source samples shared between particle models are counted once.
The independent runs estimate Monte Carlo fluctuations.
Tests check normalization, selection rules,
and four-momentum conservation in generated events, including subsequent
daughter decays.
The \exhad package~\cite{exhad} includes the numerical inputs and generator
settings. Fitting, research validation, and manuscript-figure generation
are separate from the standalone event-generation code. The working
\textsc{EventCalc-SHiP} integration is provided in its own
repository~\cite{EventCalc}.

\end{document}